\def\year{2026}\relax

\documentclass[letterpaper]{article} %
\usepackage{aaai2026}  %
\usepackage{times}  %
\usepackage{helvet}  %
\usepackage{courier}  %
\usepackage[hyphens]{url}  %
\usepackage{graphicx} %
\usepackage{natbib}  %
\usepackage{caption} %
\DeclareCaptionStyle{ruled}{labelfont=normalfont,labelsep=colon,strut=off} %
\usepackage{algorithm}
\usepackage{algorithmic}

\usepackage{adjustbox}
\usepackage{subcaption}
\usepackage{booktabs}

\usepackage{xcolor}
\newcommand{\answerYes}[1]{\textcolor{blue}{#1}} 
 
\newcommand{\answerNA}[1]{\textcolor{gray}{#1}} 

\usepackage{newfloat}
\usepackage{listings}
\floatstyle{ruled}
\newfloat{listing}{tb}{lst}{}
\floatname{listing}{Listing}
\title{A Large-Scale Study of Telegram Bots}
\author{
    Taro Tsuchiya\textsuperscript{\rm 1},
    Haoxiang Yu\textsuperscript{\rm 2},
    Tina Marjanov\textsuperscript{\rm 3}, \\
    Alice Hutchings\textsuperscript{\rm 3},
    Nicolas Christin\textsuperscript{\rm 1},
    Alejandro Cuevas\textsuperscript{\rm 4}
    }
\affiliations{
    \textsuperscript{\rm 1}Carnegie Mellon University, \\ 
    \textsuperscript{\rm 2}Tsinghua University, \\
    \textsuperscript{\rm 3}University of Cambridge, \\
    \textsuperscript{\rm 4}Princeton University \\
    ttsuchiy@andrew.cmu.edu, yu-hx21@mails.tsinghua.edu.cn, tm794@cam.ac.uk, \\
    ah793@cam.ac.uk, nicolasc@andrew.cmu.edu, aedcv@princeton.edu
}

\newif\ifdraft
\drafttrue %
\draftfalse %

\ifdraft
        \newcommand{\nicolasc}[1]{\textcolor{brown}{[[Nicolas: #1]]}}
        \newcommand{\acv}[1]{\textcolor{teal}{[[Alejandro: #1]]}}
        \newcommand{\ttsuchiy}[1]{\textcolor{red}{[[Taro: #1]]}}
        \newcommand{\yhx}[1]{\textcolor{orange}{[[Haoxiang: #1]]}}
        \newcommand{\tina}[1]{\textcolor{blue}{[[Tina: #1]]}}
        \newcommand{\alice}[1]{\textcolor{violet}{[[Alice: #1]]}}

\else
        \newcommand{\nicolasc}[1]{}
        \newcommand{\acv}[1]{}
        \newcommand{\ttsuchiy}[1]{}
        \newcommand{\yhx}[1]{}
        \newcommand{\tina}[1]{}
        \newcommand{\alice}[1]{}
\fi

\begin{document}

\maketitle

\begin{abstract}
Telegram, initially a messaging app, has evolved into a platform where users can interact with various services through programmable applications, \textit{bots}.
Bots provide a wide range of uses, from moderating groups, helping with online shopping, to even executing trades in financial markets.
However, Telegram has been increasingly associated with various illicit activities---financial scams, stolen data, non-consensual image sharing, among others, raising concerns bots may be facilitating these operations. 
This paper is the first to characterize Telegram bots at scale, through the following contributions.
First, we offer the largest \textit{general-purpose} message dataset and the first bot dataset.
Through snowball sampling from two published datasets, we uncover over 67,000 additional channels, 492 million messages, and 32,000 bots.
Second, we develop a system to automatically interact with bots in order to extract their functionality.
Third, based on their description, chat responses, and the associated channels, we classify bots into several domains. 
Fourth, we investigate the communities each bot serves, by analyzing supported languages, usage patterns (e.g., duration, reuse), and network topology. 
While our analysis discovers useful applications such as crowdsourcing, we also identify malicious bots (e.g., used for financial scams, illicit underground services) serving as payment gateways, referral systems, and malicious AI endpoints.
By exhorting the research community to look at bots as software infrastructure, this work hopes to foster further research useful to content moderators, and to help interventions against illicit activities.
\end{abstract}

\section{Introduction}
Telegram is one of the largest social apps in the world (1 billion users~\cite{telegram2025faq}) 
and an application through which people interact with the web (e.g., get news, buy goods online).
It enables individual or group chats, as well as broadcasting within channels.
In recent years, Telegram has evolved beyond a chat platform, 
and now offers comprehensive software
infrastructure: user authentication, static domains, web browsing, applications (\textit{mini apps}), and advertisements~\cite{telegram2025browser, telegram2025widget}.
Most importantly, and key to this paper, Telegram allows developers to offer
various services through \textit{bots}, programmable 
applications running within Telegram that facilitate
interaction with users through commands, messages, and inline queries. Telegram
users can directly message bots or ask them to
perform tasks (e.g., moderating messages) in a channel or group. 
Bots can also collect payments from users for goods and services, and
through traditional payment methods (e.g., 
Google or Apple Store), cryptocurrencies, Telegram's own cryptocurrency (TON)~\cite{telegram2025ton},
and Telegram's in-app currency (Telegram Stars)~\cite{telegram2025star}.

Telegram has attracted a diverse user base, ranging from 
TV show fans, students preparing for exams, to cryptocurrency traders~\cite{mozur2024telegram}.
Recent years have seen growing concerns about Telegram's association with a variety of illicit activities, leading to increased scrutiny of 
Telegram's lenient content moderation policy.
Recent evidence suggests that Telegram has become a new 
hub for data leaks and pirated software~\cite{roy2024darkgram, marjanov2025sok,marjanov2026stayin, lieber2023stolen},
fraud~\cite{gao2020tracking,cernera2023token, bijmans2021catching},
non-consensual image abuse distribution~\cite{burgess2020telegram,semenzin2020use}, money
laundering~\cite{gebrekidan2023thescammer},
propaganda~\cite{kireev2025characterizing, hanley2024partial}, and 
extremist groups~\cite{mozur2024telegram}. 
Anecdotal evidence suggests
that bots are crucial in scaling these illicit
operations because they allow operators to automate interactions with users and
content.

However, little is known about the services bots provide since past work has largely focused on messages, channels, and groups~\cite{pushshift,tgdataset,guo2024beyond,blas2025unearthing,kireev2025telegram,gangopadhyay2025telescope}.
In addition, Telegram does not have a centralized bot repository, making it difficult to enumerate these bots.
To fill this gap, we offer the following contributions.

\noindent \textbf{1.} We provide the largest \textit{general purpose}
message dataset and the first bot dataset on Telegram. 
We update existing public datasets, extract additional 
channels and messages through snowball sampling, and extract the list of bots.
Our dataset contains approximately 
106,000 channels and 809 million messages, including 67,000 channels and 492 million messages that had not been reported before, 32,000 bots, and 9 million links between bots/channels/users.

\noindent \textbf{2.}  We
develop a novel system to autonomously send commands to each bot and record its
responses. 
This is necessary to understand bot functionality, since the bot description is often insufficient.

\noindent \textbf{3}. To understand what services bots offer, we classify the bots
into several
domains and functionalities based on our largest dataset above. We label all bots using a combination of manual,
keyword, and Large Language Models (LLMs) analyses. While most bots are benign, we find an alarming number of bots with illicit operations such as fraud (4\%) and underground (5\%) (e.g., nudification apps, unauthorized access to paid content, and stolen data). 
Those bots often process payment, manage referrals, give access to ill-gotten digital goods, or host malicious AI endpoints. 

\noindent \textbf{4}. We analyze the communities around bots to understand the bot usage. 
Despite descriptions being in English, many bots
are used in non-English communities. While Russian is the most common language for
most categories, English is predominant among finance bots. 
Finance bots are the most prevalent, but are often short-lived, 
reused, and frequently flagged by Telegram as scam. 
Ideology bots are often designed for a specific, close-knit
community, whereas utility bots appeal to broader sparse communities.

Based on our analysis, we argue that bots have become a core infrastructure for facilitating illicit activities on Telegram, for instance, serving as payment processors, referral systems, or malicious AI endpoints. 
We suggest that Telegram and law enforcement focus on bot moderation as a potential point of disruption.
We offer four practical recommendations: 1) looking at the bot command list and messages, 2) focusing on specific domains based on language, 3) grouping similar bots, and 4) identifying related channels through links. 
Our findings highlight the need to study Telegram ecosystem beyond messages
and groups. Our dataset can serve as a valuable resource enabling further
research. 
Following Ethical Statement, the dataset and the code are available at \url{https://zenodo.org/records/17281308} and \url{https://github.com/taro-tsuchiya/TelegramBots}, respectively. 

\section{Background}
\label{sec:related}
We start with an overview of Telegram, discuss related work,
and compare our data with previously released datasets.

\subsection{Telegram Overview}
\label{subsec:telegram_overview}
Since its 2013 launch, the Telegram messaging app has
positioned itself as a secure and privacy-focused alternative to its competitors. 
It has grown to 
over 1 billion active users~\cite{telegram2025faq}, and has become a key
messaging channel for various communities, businesses, news organizations, and even
heads of state. However, Telegram has also been criticized for its lax content
moderation policies, which have allowed it to become a hub for various illicit
activities~\cite{mozur2024telegram}.

Today, Telegram is a platform which offers various features for users
and developers, ranging from group messaging to advanced web services. 
Telegram supports two modes of group communication: \textit{Channels} (one to many, unlimited number of members) and \textit{groups} (many to many, up to 20,000 members).  
Furthermore, websites can authenticate users
with the Telegram widget~\cite{telegram2025widget}, and even verify
ID with the Telegram Passport program~\cite{telegram2025passport}.
Telegram also supports web-based applications (``mini-apps,''
e.g., games, e-commerce) that can be launched within Telegram or a
fully-fledged in-app browser~\cite{telegram2025browser}. Users do
not need to leave Telegram to access mini apps or websites. Telegram
further offers a variety of payment integration options: traditional
credit cards, cryptocurrencies, the newly launched ``Telegram
Stars''~\cite{telegram2025star}, an in-app currency to facilitate
transactions between Telegram users, and even its own blockchain---The
Open Network (TON)~\cite{telegram2025ton}---originally developed for
cryptocurrency payments into the Telegram ecosystem. Users can
use TON and the associated Toncoin cryptocurrency to purchase goods,
services, and even advertisement space within Telegram.

Telegram \textit{bots} are the primary way to offer goods
and services programmatically within Telegram. Bots provide a text-based
interface to users, allowing them to interact with them through commands and
other inputs. For example, a user can provide an URL to initiate a
download or provide an image for processing. Bots can
also request payments and manage memberships, all without leaving Telegram. To
create a bot, developers first register a new bot through Telegram's
``BotFather,'' a special bot that helps users manage their bots. They
\textit{must} choose a username that ends with ``\texttt{bot},'' (case insensitive), 
and obtain an API token. 
Developers then write a script in any programming language (e.g., JavaScript, Python), 
run it on their own device/server or cloud, and connect it to Telegram through the Telegram Bot
API and their API token. Telegram requires developers to support two global
commands to ensure uniform user experience: \texttt{/start} (to launch the
description and welcome message) and \texttt{/help} (to list all functions)~\cite{telegram2025features}.  

\subsection{Related Work}
\label{subsec:related_work}
Because of Telegram's lenient content moderation policy, Telegram has become a
hub for problematic content such as
propaganda~\cite{kireev2025characterizing, hanley2024partial},
disinformation~\cite{ng2024exploratory},
conspiracies~\cite{steffen2025more, imperati2025conspiracy}, cybercrime~\cite{guo2024beyond,
roy2024darkgram, marjanov2026stayin}, and
phishing~\cite{gao2020tracking,cernera2023token,bijmans2021catching}.
Cryptocurrency and blockchain communities are also active on
Telegram~\cite{pushshift}, and several 
studies document cryptocurrency price
manipulation coordination attempts~\cite{nizzoli2020charting,xu2019anatomy,mirtaheri2021identifying}. 

Despite the critical role that bots play in the Telegram ecosystem, only a
handful of studies investigate or even mention them. \citet{roy2024darkgram} characterize the use of bots for cybercrime in their
Appendix, discussing payment processing, content
distribution, and channel expansion. 
\citet{ricaldi2025uncovering} show that Telegram bots can be used for gaining customers' trust in underground markets (e.g., to automate the order process). 
\citet{alrhmoun2024automating} study Telegram bots in channels supporting
the Islamic State, identifying two roles: content distribution and group
management. 
\citet{perlo2025topic} show the prevalent use of Telegram bots; nearly 90\% of their groups include bots. 
Other papers mention the use of Telegram bots for sharing non-consensual images~\cite{semenzin2020use, franco2024characterizing}, controlling IoT devices~\cite{de2016chatting}, or managing groups~\cite{nikkhah2018telegram}.
Recent web articles describe the use of bots to help cybercriminals communicate with
malware-infected devices~\cite{nigam2018user}, support phishing operations~\cite{buyukkaya2024onnx},
non-consensual image abuse~\cite{burgess2020telegram}, or recruit members for
extremist groups~\cite{mozur2024telegram}. However, our 
paper is the first to characterize bot roles and functionality
at scale, for both legitimate and malicious uses.

\subsection{Comparison with Previous Datasets}
\label{subsec:dataset_comparison}
We provide 1) the largest general-purpose Telegram
message dataset, and 2) the first bot dataset, including bot/channel
descriptions and linkage, and our interactions with bots. We review past
Telegram datasets and compare them to our work. We limit our comparison
to those 1) are publicly hosted and accessible, 2) documented, 
and 3) contain at least 1~million messages.
Table~\ref{tab:related-dataset} summarizes the seven existing such datasets and ours. 
Some datasets cater to specific topics such as
politics~\cite{pushshift,blas2025unearthing}, propaganda~\cite{kireev2025telegram}, war~\cite{bawa2025telegram}, and underground
markets~\cite{guo2024beyond,marjanov2026stayin}, while others (including ours) are general
purpose~\cite{tgdataset,gangopadhyay2025telescope}. Most works start from seed 
channels (through a website, e.g., TGStat or
keyword search) and use snowball sampling to discover more channels 
(as we do in \S\ref{subsec:collect_channels_messages}). 
As Table~\ref{tab:related-dataset} shows, our data include the largest 
general-purpose 492 million message dataset, and a unique bot dataset.

\begin{table*}[]
    \begin{adjustbox}{width=\textwidth,center}
    \begin{tabular}{@{}lllllll@{}}
    \toprule
    Author             & Year / Venue          & Nr of channels & Nr of messages
    & Collection date & Topic                     & Snowball \\ \midrule
    \citet{pushshift}  & 2020, ICWSM           & 30K
    & 317M           & $\sim$Oct 2019  & Politics/Crypto. & Yes      \\
    \citet{tgdataset}    & 2025, KDD             & 120K
    & 400M           & $\sim$Jul 2022  & General                   & Yes      \\
    \citet{guo2024beyond}          & 2025, SIGMETRICS      & 71K
    & 200M           & $\sim$Feb 2024  & Underground          & Yes      \\
    \citet{blas2025unearthing}         & 2025, WWW & 43K
    & 1B             & $\sim$Feb 2025  & Politics                  & Yes      \\
    \citet{kireev2025telegram}       & 2025, ICWSM           & 13
    & 13M            & $\sim$Oct 2023  & Propaganda                & No       \\
    \citet{bawa2025telegram}       & 2025, ICWSM           & 518
    & 5M            & $\sim$Mar 2023  & War                & No       \\
    \citet{gangopadhyay2025telescope} & 2025, ICWSM
    & 71K            & 120M           & $\sim$Oct 2024  & General
    & Yes      \\
    \citet{marjanov2026stayin} & 2026, USENIX Sec. & 1,521        & 14M    & $\sim$Aug 2025  & Underground & Yes  \\ \midrule
    \textit{This paper} & \textit{2026, ICWSM}                  & \textit{106K (68K new)}      & \textit{809M (492M new)}
                        & \textit{$\sim$Aug 2024}  & \textit{General}                   & \textit{Yes}      \\ \bottomrule
    \end{tabular}
    \end{adjustbox} 
    \caption{Comparison of publicly available Telegram datasets.}
    \label{tab:related-dataset}
\end{table*}
\section{Data Collection}
\label{sec:methodology}
We explain how we collect Telegram channels\footnote{For our analysis, we do not distinguish channels and groups and refer both as channels. 82.2\% are channels and 18.8\% are groups.}, messages, and bots.

\subsection{Collecting Channels and Messages}
\label{subsec:collect_channels_messages}

\begin{figure}
	\centering
	\includegraphics[width=\linewidth]{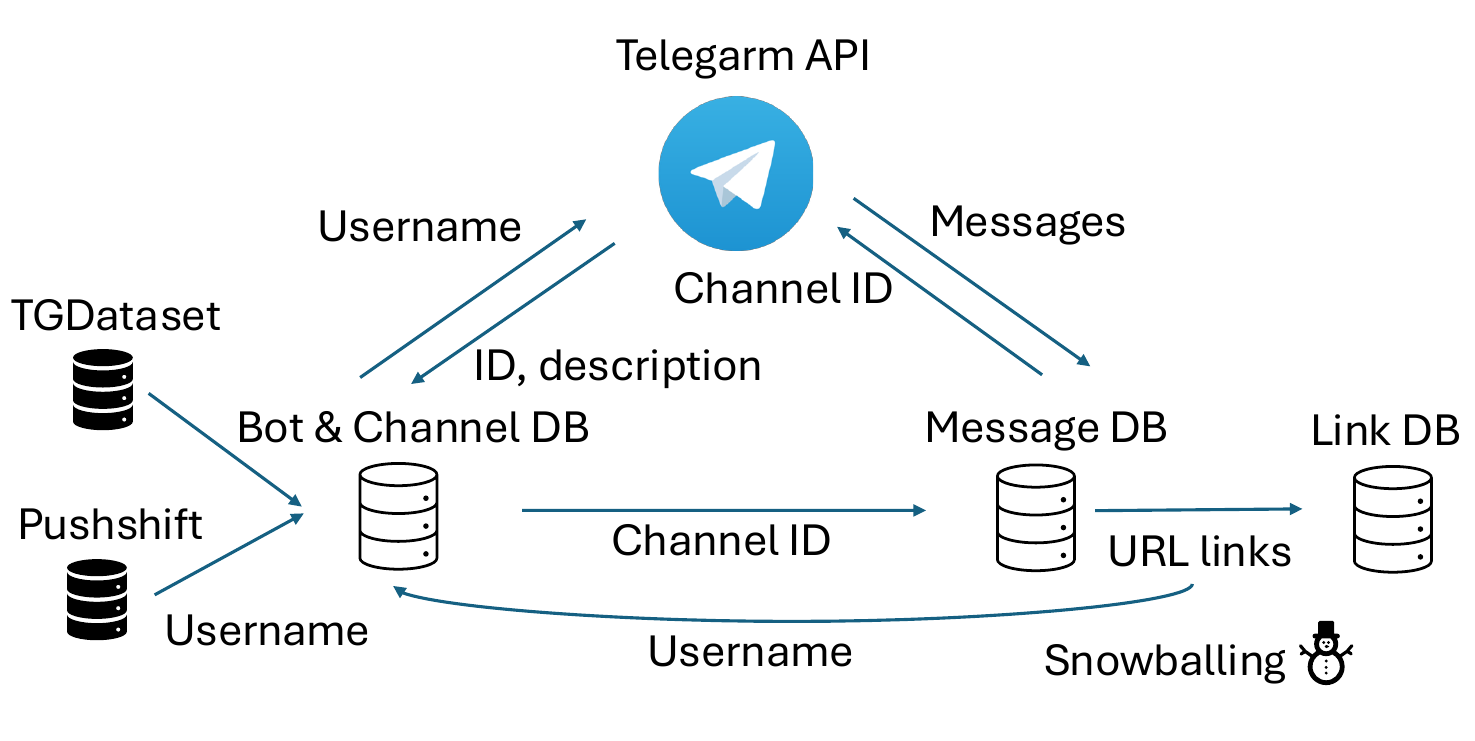}
	\caption{Data collection pipeline}
	\label{fig:data_collection_pipeline}
\end{figure}

\begin{figure}
	\centering
	\includegraphics[width=\linewidth]{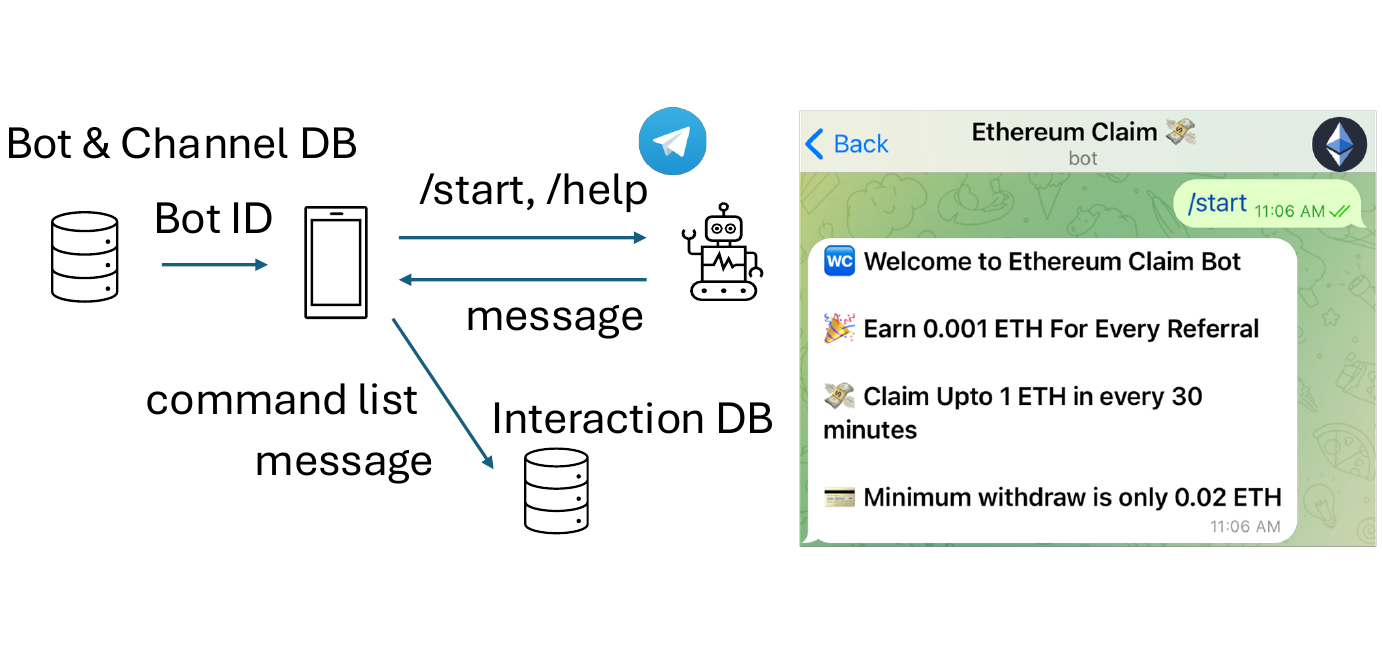}
	\caption{Bot interaction pipeline}
	\label{fig:bot_interactionpipeline}
\end{figure}

We show our data collection pipeline in Figure~\ref{fig:data_collection_pipeline}.
We start our snowball sampling from the channels from the Pushshift~\cite{pushshift} and TGDataset~\cite{tgdataset} datasets. As 
Table~\ref{tab:related-dataset} illustrates, Pushshift~\cite{pushshift} features over 30,000
channels and primarily focuses on right-wing or cryptocurrency-related channels,
whereas TGDataset~\cite{tgdataset} includes nearly 120,000 channels and covers a
broader variety of topics (e.g., US news, entertainment). Combining both datasets
provides a general overview of the Telegram ecosystem. As both
datasets are old (last collected in Oct. 2019 and Jul. 2022, respectively),
we begin by updating messages from those channels. 

Telegram does not have algorithms to recommend bot/channels, 
so their information is often passed through URL links in messages. 
We extract 
channel names from URLs using regular expressions that match patterns such as
\texttt{telegram}, \texttt{tg}, \texttt{t.me/}, and others. After verifying the
existence of those channels through the API, we conduct snowball
sampling---iteratively collecting new channels through URL links from new messages. For each channel, we also collect its description.
Besides discovering new channels, we also continuously collect new messages
from the channels already collected. Notably, Telegram lets us get 
public channel metadata and messages passively, i.e., without joining the channel.
We do not join any private
channels or collect any media files (e.g., images, videos, documents).

Our collection ran continuously from May 19th until Aug. 18th,
2024, resulting in a dataset of 105,970 channels and 809,481,087 messages. 
We discovered 67,868 new channels (not in Pushshift and TGDataset) and 492,256,372 messages (not in Pushshift).\footnote{TGDataset stores messages differently
from Pushshift or us, so we could not directly compare against it.} We extract
9,041,103 URL links to other channels/bots/users, far more than previous efforts 
(2.7 million links by \citet{gangopadhyay2025telescope}).

\subsection{Bot Discovery}

\noindent \textbf{Bots Collected from Messages}. Bot information is
typically passed through user links (e.g.,
\texttt{t.me/USERNAME})~\cite{telegram2025features}. For bots, \texttt{USERNAME} 
\emph{must} end with (case-insensitive) ``\texttt{bot}'' (see \S\ref{subsec:telegram_overview}). 
We look at channel messages\footnote{We might miss bots that are not mentioned in messages but only in channel descriptions.} and extract all user links and exclude channels/groups based on API response.
For the remaining user links, we extract the usernames ending with ``\texttt{bot}.''
Two of the authors checked
100 random samples and found no
obvious false positives---i.e., regular user handles serendipitously ending with ``\texttt{bot}.'' 

\noindent \textbf{Bots Collected from Third-Party Directories}. Because
Telegram does not have a centralized bot directory, users also rely on
third-party websites that list bots and channels. To increase our coverage, we also extract the bots from 
\url{telegramic.org/bots}, \url{tgbots.io/}, \url{telegrambotlist.com}, \url{tgdr.io}, and \url{telega.io}. 

Combining both sources yields a total of 32,071 bots (23,886 (74.5\%) through 1,147,674 URL links across 17,455 channels; and 8,185 (25.5\%) through third-party directories). We verify that each 
bot is active and not deleted, and collect its description through the Telegram APIs: \texttt{userFull} and
\texttt{botInfo}. The description includes the ``\texttt{about}'' field in the bot
profile, the description of the bot (at start time), and the list
of URLs displayed in the bot profile. 
20,416 (63.7\%) bots have non-empty
descriptions.

\subsection{Extracting Bot Functionality}
\label{subsec:bot_interaction}
While some bots may use their description to detail their functions in the description, many bots instead rely on 
1) external documentation or 2) interactions with users. The former is more
common for bots that carry out routine tasks (e.g., moderation) within groups---we can extract this information from messages. To cover the latter case, we interact with bots. 
Figure~\ref{fig:bot_interactionpipeline} illustrates the interaction pipeline, along with one example of a bot interaction.
We launch the Telegram client (using the official API) and issue two required 
commands (\texttt{/start}, \texttt{/help} (see \S\ref{subsec:telegram_overview})) to all bots in our dataset, and record responses within a 5$s$ timeout.
Bots whose script is not running return no response. 
In addition, not all bots adhere to this standard. In total, 7,375 bots responded to either \texttt{/start} or \texttt{/help} commands. (4,959
bots responded to \texttt{/start} and 5,502 bots responded to \texttt{/help}.)
When the bot is flagged by Telegram as a scam or fake account, it will respond
with a warning message;\footnote{``Warning: Many users reported this account as a scam or a fake account. Please
be careful, especially if it asks you for money.''} we found 1,331 such bots.
This number is a lower bound, as scam bots could have been deleted
prior to our data collection. Last, 
9,197 bots returned a command list upon request. 67\% support \texttt{start}; 28\%
, \texttt{help}, 7\%, \texttt{balance}; 7\%, \texttt{settings}; and 6\% feature 
\texttt{menu} commands.
\section{Characterizing Bots}
\label{sec:categorization}
To understand the bots' services, we classify all 32,071 bots into the domains they
operate in and the functionalities they provide. 

\subsection{Usage Domains}
\label{subsec:domains}
We first illustrate how we categorize the domains, and then show
the classification results.

\subsubsection{Methodology}
\label{subsubsec:domains_methodology}
One author, who is an active Telegram bot user, initially came up with
21 domains based on the Telegram channel topics (TGDataset) and the websites
with the list of bots (\S\ref{sec:methodology}). 
As a test run, two authors independently annotated 100 test samples,
discussed disagreements, and merged less common categories.

We ended up with nine categories:
\begin{enumerate}
    \item \textbf{Admin Tools} (AT): Bots that manage groups on behalf of
    owners, such as membership management, question answering, content
    moderation, and group statistics. 
    \item \textbf{Content \& Media} (CM): Bots that help distribute or collect
    educational and training materials, streaming (music, movies, TV series),
    and news media. 
    \item \textbf{Ideology} (ID): Bots used for political campaigns,
    social movements, or religion-related purposes.
    \item \textbf{Finance} (FN): Bots that provide access to financial services,
    such as online wallets, trading, airdrops, mining, or providing financial
    information.
    \item \textbf{Shopping} (SP): Bots that facilitate online shopping,
    including selling and buying products, collecting reviews, providing
    customer service, and product search.
    \item \textbf{Social \& Gaming} (SG): Bots that facilitate online
    interactions (e.g., chatting for fun, dating, games, and
    gambling).
    \item \textbf{Underground} (UG): Bots that support underground operations,
    such as cybercrime (e.g., hacking, stolen data, phishing) and adult content.
    \item \textbf{Utility} (UT): Bots that provide tools or functions to
    individual users (not groups), such as using LLM endpoints, developer tools,
    web search, photo \& video management, health \& fitness management, and QR
    code generation. 
    \item \textbf{Other}: Bots that do not belong to the categories above or are
    unknown given the input.
\end{enumerate}

Using this categorization, two authors independently annotated another random sample of 100 bots based on their
usernames, descriptions, interactions, and the messages
containing links to the bots. We treat each source of information equally, but look at the messages that may contain 
other contexts with caution. 
We choose only one category for each bot. If the
bot operates in multiple domains, we follow the instructions specified in the
prompt (see Appendix~\ref{sec:appendix_prompt}). 
For instance, we classify ``selling Netflix accounts'' as Underground
rather than Shopping or Content \& Media. If not specified, the coders chose the
most prominent category. 

We only include messages that mention one bot, to 1) 
exclude messages enumerating many bots for advertisement, and 2) 
avoid referring to other bots. We only use the first 50 messages or a maximum of
2,000 characters to reduce the amount of text (for both LLMs and human
annotators). We manually find that those thresholds are
sufficient to capture the context. We provide both the original text and
the English translation (using Google Translate) to help with human
annotation. Coders are allowed to search keywords online, but not allowed to 1)
search or interact with the bot directly or 2) ask LLMs for help.  

The interrater agreement, Cohen's $\kappa$~\cite{cohen1960coefficient}, for the two
coders is 0.67, which is considered substantial agreement. Most of the
discrepancies arise from 1) speculation based on the
keywords in the username (e.g., is the keyword ``ETH'' sufficient to select
Finance?), 2) ambiguity regarding the bot's target audience (e.g., is a
Q\&A bot intended for use by administrators or individual users?), 3) lack
of domain knowledge (e.g., a clothing brand, an anime character), or 4) less
commonly, ambiguity in the prompt. We discuss the disagreements one by one,
reach a consensus, and slightly modify our codebook to clarify ambiguity.

We next use OpenAI LLMs to scale up the annotation to \textit{all} bots. To verify the
accuracy of LLMs, we run gpt-4o (2024-08-06) and gpt-4o-mini (2024-07-18) on the same 100
samples first and compare the results with human
annotation (consensus). We do not translate any text and directly feed the
original text to LLMs. For model parameters, we set \textit{temperature} to 0
and \textit{top\_p} to 1 to reduce randomness and improve the reproducibility
(default values for other parameters). The agreement between human consensus and LLMs is 0.73 for gpt-4o and 0.70 for gpt-4o mini.
After a manual inspection on all samples, most of the disagreements come from
reasons similar to those observed with human annotators---bot description ambiguity,
e.g., ``coin'' can be both in Finance or Social \& Gaming depending on the
context. We rarely find cases where LLMs make mistakes due to misunderstanding
the context, but rather sometimes correct humans' mistakes. Between two LLMs,
the gpt-4o model appears to be slightly more conservative than the gpt-4o-mini
model (e.g., uses Other more often, rarely makes judgment just based on
username), while the differences appear to be minimal. We primarily use gpt-4o classifications for further analysis.
Appendix~\ref{sec:appendix_annotation} provides the annotation results (e.g., domain distribution) for two human annotators and two LLMs.

\subsubsection{Results}
\label{subsubsec:domains_results}
\begin{figure}
    \centering
    \includegraphics[width=1\linewidth]{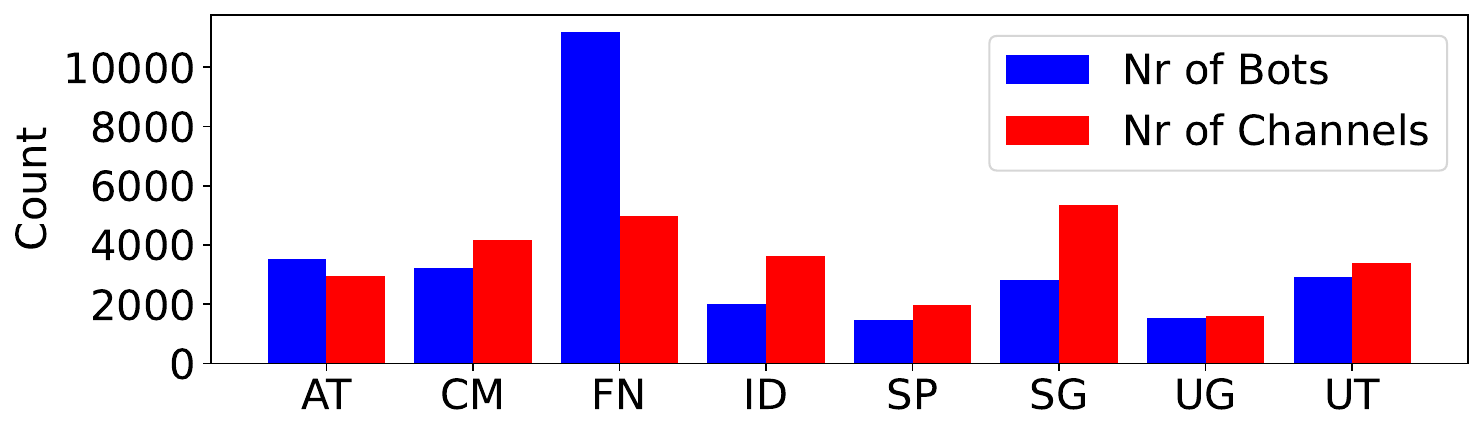}
    \caption{Number of bots and channels per category.}
    \label{fig:domain_dist}
\end{figure}

We apply the same prompt for the entire sample and run two LLMs. We also ask
them to produce a one-sentence summary so that we can review the categorization.
Figure~\ref{fig:domain_dist} shows the number of bots per category. We have 34\%
(n=11,206) of bots in Finance, 11\% (n=3,517) in Admin Tools, 10\% (n=3,222) in Content \&
Media, 9\% (n=2,920) in Utility, 9\% (n=2,818) in Social \& Gaming, 6\%
(n=2,004) in Ideology , 5\% (n=1,539) in Underground, 5\% (n=1,451) in Shopping,
and 11\% (n=3,394) in Other. We also aggregate the
number of unique channels that mention those bots for each category. As
Figure~\ref{fig:domain_dist} shows, Finance has a disproportionately high number
of bots compared to its number of channels, while Social \& Gaming shows the
opposite. This indicates that Finance bots are distributed by a smaller set of
channels. Indeed, when we look at the top ten channels that mention the most
bots, nine of them are advertisement channels that are dedicated to the
promotion of Finance bots (i.e., 70-100\% are Finance bots). Those top ten
channels discovered 4,674 Finance bots (41.7\% of all Finance bots).

\subsection{Analyzing Functionality}
\label{subsec:functionality}
Through the manual bot annotation process in \S\ref{subsubsec:domains_methodology}, three common sets of capabilities emerged
across bots: payment processing, referral management, and input collection. 
Additionally, we observed a substantial number of bots using artificial
intelligence to provide services. We refer to these
functionalities as Payment, Referral, Crowdsourcing, and AI.

\subsubsection{Payment}
\label{subsubsec:payment} 
We first examine whether bots process payments. To do this, we use a keyword
matching approach, whereby we classify the payment functionality based on the
command list that the bot provides. Initially, we attempted to use LLMs to infer
functionalities, but they often failed to distinguish between the bot's
functionalities (i.e., what it actually does) and the context in which the bot
is used (e.g., the airdrop bot does not always process payment, but just
helps expand channels). The keyword matching approach reduces false positives
at the expense of being more conservative, thus providing a lower bound. We
select the list of keywords based on 1) expert knowledge and 2) manual
inspection of the frequent bot commands. We translate all commands
and their descriptions to English using the Google Translate API. We
follow the same process for Referral and Crowdsourcing functionalities.

We select the following keywords: ``pay,'' ``payment,'' ``purchase,'' ``buy,''
``sell,'' ``deposit,'' ``withdraw,'' and ``setwallet.'' To verify the accuracy of keyword matching, we manually checked 100
random command name and descriptions and found no false positives. In total, 9.4\%
(864 out of 9,197 bots with a command list) have payment-related commands. 
We manually group the command variants (e.g., \texttt{withdrawal},
\texttt{quickwithdraw}) to the base command (e.g., \texttt{withdraw}). The most
common commands are to \texttt{withdraw} (n=474), \texttt{wallet} (n=369),
\texttt{deposit} (n=206), \texttt{pay} (n=202), and \texttt{buy} (n=114).
Users are more likely to withdraw than to deposit because they can cash out
without initial deposit (e.g., airdrops, referrals). Bots also set or
connect wallets, buy service credits, create payment links, invest, and upgrade to VIP
memberships. We could not identify the type of
payment method (e.g., crypto, credit card), since figuring this out 
typically involves additional interaction.

\subsubsection{Referral}
\label{subsubsec:referral}
We examine whether the bot produces referral
links or asks users to invite friends. We follow the same strategy as Payment
and select the following keywords: ``refer,'' ``referral,'' and ``invite'' for matching. We explicitly remove ``references'' because it is often used
in different contexts (e.g., to reference a book). We only focus on the
referral aspects, so we do not count the ``gateway'' bot that simply asks users
to join the corresponding channel (e.g., \texttt{join} command). From randomly
chosen 100 commands, we find no false positives (by only looking at the command descriptions). In total, 10.4\% (n=954)
of bots have referral-related commands. Those bots typically issue customized
referral links to users and ask them to invite their friends: \texttt{referral}
(n=638), \texttt{invite} (n=296). Based on the command description, they
also give rewards (``earn credit''). The form of rewards is either free
cryptocurrencies, tokens (e.g., TON), or VIP subscriptions.

\subsubsection{Crowdsourcing}
\label{subsubsec:crowdsource}
We investigate how the bot collects textual information or files from a group of
users. Likewise, we choose the following keywords: ``upload,'' ``submit,''
``report'' to match the command list. We exclude \textit{report\textbf{s}}
(plural) command because it is often used to produce statistics. We manually
find 11 false positives out of a random 100 commands, mostly
\texttt{report} command, used for statistics. We ultimately decided against excluding ``report''
as it is generally used for accepting inputs from users. In total, we have
1.2\% (n=113) bots with crowdsourcing-related commands: \texttt{report} (n=64),
\texttt{upload} (n=17), and \texttt{submit} (n=12).

\subsubsection{AI}
\label{subsubsec:ai}
We finally investigate whether bots provides a service based on artificial
intelligence (AI). Bots do not necessarily have a specific set of commands for
AI-powered services, and the definition of ``AI-powered'' service is vague, so
we perform the keyword matching on the bot's (translated) description. We generate 45 AI-related
keywords, ranging from basic machine learning, such as ``ml,'' ``ocr,''
``translation,'' to the recent generative models such as ``chatgpt.''
We do not allow partial matches because the description often contains a large
set of words (e.g., ``airdrop'' matches ``ai''). We manually check 100 random bots to
verify the accuracy of keyword matching, and find that the keyword ``transformer''
is sometimes used as a robot but not the AI model, so we exclude it. 607 bots
have AI keywords in their description. The most common matched keywords
are general descriptions of AI:  ``ai'' (n=363), ``artificial-intelligence''
(n=96), ``translation'' (n=42). We also observe many bots using the
generative model: ``chatgpt'' (n=69), ``gpt'' (n=58), or image-generation:
``midjourney'' (n=20), and ``dall-e'' (n=15).

\subsection{Domains and Functionality}
\label{subsec:domains_functionality}
\begin{figure}
    \centering
    \includegraphics[width=1\linewidth]{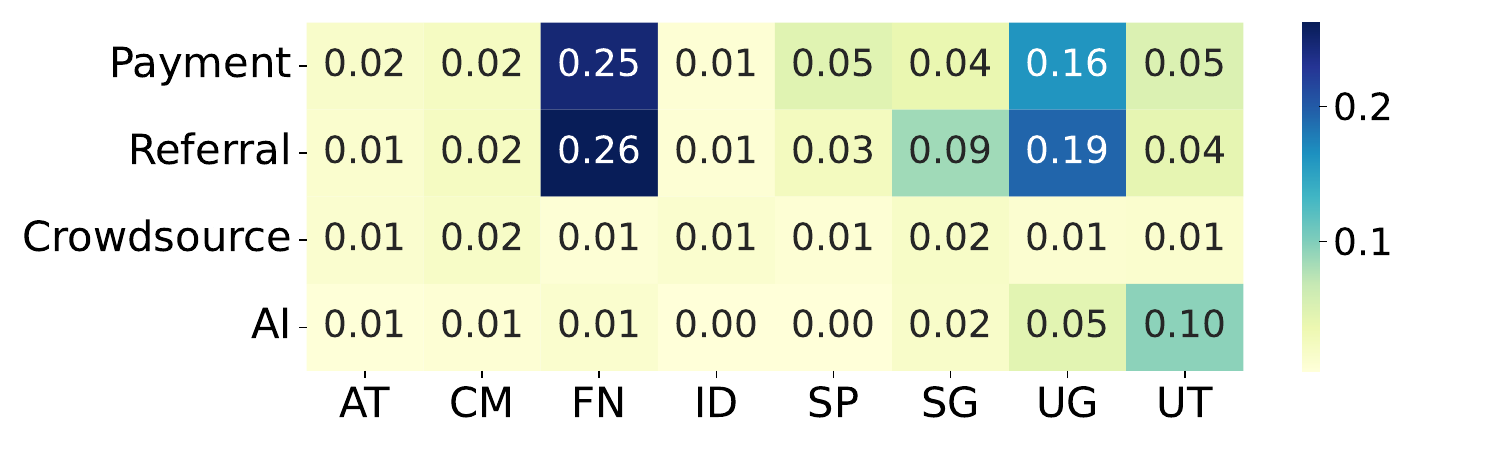}
    \caption{Distribution of bot domains and functionalities, e.g., 25\% of Finance bots (with a command list) have a payment functionality. }
    \label{fig:domain_functionality}
\end{figure}

We quantitatively show the relationship between the domains and functionalities.
Figure~\ref{fig:domain_functionality} shows the distribution of bot domains
(x-axis) and functionalities (y-axis). We divide by the total number of bots in
each domain by the number of bots \textit{with} command list (all bots for AI).
Each bot can have multiple functionalities. 
Finance (FN) and Underground (UG) are significantly more likely to use bots as a
payment processor or a referral system compared to other domains. While Shopping
(SP) bots have a relatively low payment functionality given their nature, those
bots still support online shopping through product search (e.g., catalog, price) 
or customer service (Q\&A, reviews). Content \& Media (CM) and Social \&
Gaming (SG), the two categories that often connects users, 
have relatively higher rates for crowdsourcing. 
In terms of AI usage, we find that Underground
(UG) and Utility (UT) are the most prominent. The next section explains why
certain domains predominantly use specific functionalities, and introduces some
case studies for both benign and malicious use cases.

\subsection{Benign Uses}
\label{subsec:benign_use}
We identify a variety of benign uses. For example, crowdsourcing bots mostly
support legitimate use cases. First, there are bots that accept user input from
various geographical locations. For example, we find a bot in Tbilisi that
collects information about a polluted area in the city, or a bot in Greek
refugee camps that aggregates the volunteer information or field status. 
Second, bots can assemble knowledge from the crowd,
such as creating a public library---collecting and distributing books. Third,
content moderation bots accept user reports on problematic users or messages.

There are several benign apps that use AI. Bots often perform basic tasks such
as translation, OCR (e.g., extracting texts from photos), photo editing, or
internet search, supporting the high use of AI in Utility bots. In addition,
some bots utilize generative AI models to converse with users as a friend (e.g.,
for fun or language learning) or a supporter (e.g., to provide emotional or
fitness support). Interestingly, there is a bot that allows users to use AI
anonymously (and use cryptocurrency as a payment), which could be used for
circumventing censorship. In Finance, we frequently observe some bots that use
AI for trading (analyzing social media or assets).%

\subsection{Malicious, Illicit, and Exploitative Uses}
\label{subsec:malicious_use}
We next identify various cases in which bots help facilitate malicious
activities (e.g., scams and fraud), illicit (e.g., selling cybercriminal goods,
piracy), and exploitative (e.g., undressing apps). The following
analyses are based on sampling of 100 bots with Finance \& scam labels and 100
bots categorized with the Underground domain.

\noindent \textbf{Scams and Fraud}. Bots used for scams were the most prominent.
Telegram currently has a label for bots that are suspected of being scams, which
is visible when users try to start a conversation with the bot (\S\ref{sec:methodology}). 
Around 10\% of the
Finance bots in our dataset had been assigned a scam warning, significantly higher than the
1\% average in non-Finance domains. We also find that Finance \textit{scam} bots have
exceedingly higher payment and referral functionality (71\% and 65\% respectively),
suggesting that they often ask or offer payment to users and employ tactics to
attract more users. To understand the nature of those bots, we investigate their
types and tactics.

A substantial number of bots offered airdrops or giveaways. Those bots typically
use ``free'' as a keyword, and solicit users to 1) join a channel, 2) invite
friends, or 3) watch advertisements. Some bots use the term ``mining,'' which is
not necessarily related to mining blockchain with computational resources, but
as a way to earn tokens in online games. The second prominent case was
investment bots that promise a high return by depositing or adding liquidity to
the pool. For instance, one bot offers users a chance to earn tokens without
investment, but can earn faster by investing more. Most bots are
related to cryptocurrencies such as BTC, BSC, and TRX and sometimes ask users to
provide their blockchain addresses. 
The results verify the high payment and referral functionalities of Finance.
Lastly, we also identify non-financial scams
such as referral schemes or offering free items (in Underground). 

\noindent \textbf{Illicit Goods and Services}. Many bots enable underground
markets offering illegal goods or services, such as stolen data (e.g., ``combo''
lists (URL, login, password), credit cards, breached
databases), drugs, or doxing (e.g., ``bombing'' texts for annoyance). For
example, users can use bots to access inventory of leaked databases containing
personal information through lookup commands such as \texttt{name} and
\texttt{phone}. Another common category of goods and services were those
associated with artificial engagement (e.g., views, likes, members). We also
identified bots allow access to unauthorized content, likely violating the terms
of service of the original providers. For instance, there are bots that generate
phone numbers, giving access to streaming services (e.g., Netflix) for free or
referral points, premium accounts of various apps (e.g., VPN services, online
shopping, social media), or reselling accounts. These bots typically monetized
their services through direct in-app payment or membership plans, and employed a
variety of referral tactics to attract users. These results show an increasing
trend of illicit activities which were commonly offered through Tor hidden
services, but are now facilitated through Telegram in a more user-friendly and
mobile-first manner.

\noindent \textbf{Exploitative Content}. Many bots facilitate the search,
distribution, or generation of adult content. More alarmingly, the majority of 
those bots offer services to create deepfake images, 
undressing pictures, or swapping faces, which aligns with the surge of nudification
websites~\cite{han2025characterizing,gibson2025analyzing} and non-consensual
image sharing channels on Telegram~\cite{semenzin2020use}. While these bots
could be used with the consent of the image subject, some of them are advertised
for potentially non-consensual purposes. Some bots provide a
disclaimer that users must be over 18 or adults to use the application, although
they may not have mechanism to verify the age. 
\section{Bot Usage}
\label{sec:analysis}
This section analyzes channels surrounding each bot to understand the bot usage: supported languages, duration, reuse, and topology.

\subsection{Languages}
\label{subsec:languages}
Language is an important aspect of bot usage as it can reflect the target
geographic audience. We run the language detection (using langdetect library~\cite{danilk2025langdetect}) on both 1) the bot
description and 2) the messages that mention the bot. 
Our data contains short text, abbreviations, and slang, which make language detection challenging. 
To address this, we remove 
URLs and emojis before detection and perform several rounds of
adjustments where we manually inspect 200 classifications and add/subtract a
bias in detection probability to the languages that are often
under-/overclassified until no significant improvements can be made. 
For the bot description, 42\% of bots are in English, 26\% in Russian, 11\% in
Farsi, and 6\% in Arabic. For the messages that mention bots, we run the
detection for all deduplicated messages that are associated with each bot, and
take the most frequent language. 30\% are in Russian, 30\% in English, 13\% in
Farsi, and 8\% in Arabic. 
This suggests that bots that have descriptions in English are often used in non-English speaking
communities. We also confirm that bots whose description is in Russian, Farsi, and
Arabic are mostly used by the same language communities. This result highlights the
importance of looking at the contexts (i.e., collecting messages) to understand
the bot usage. 

\begin{figure}
    \centering
    \includegraphics[width=1\linewidth]{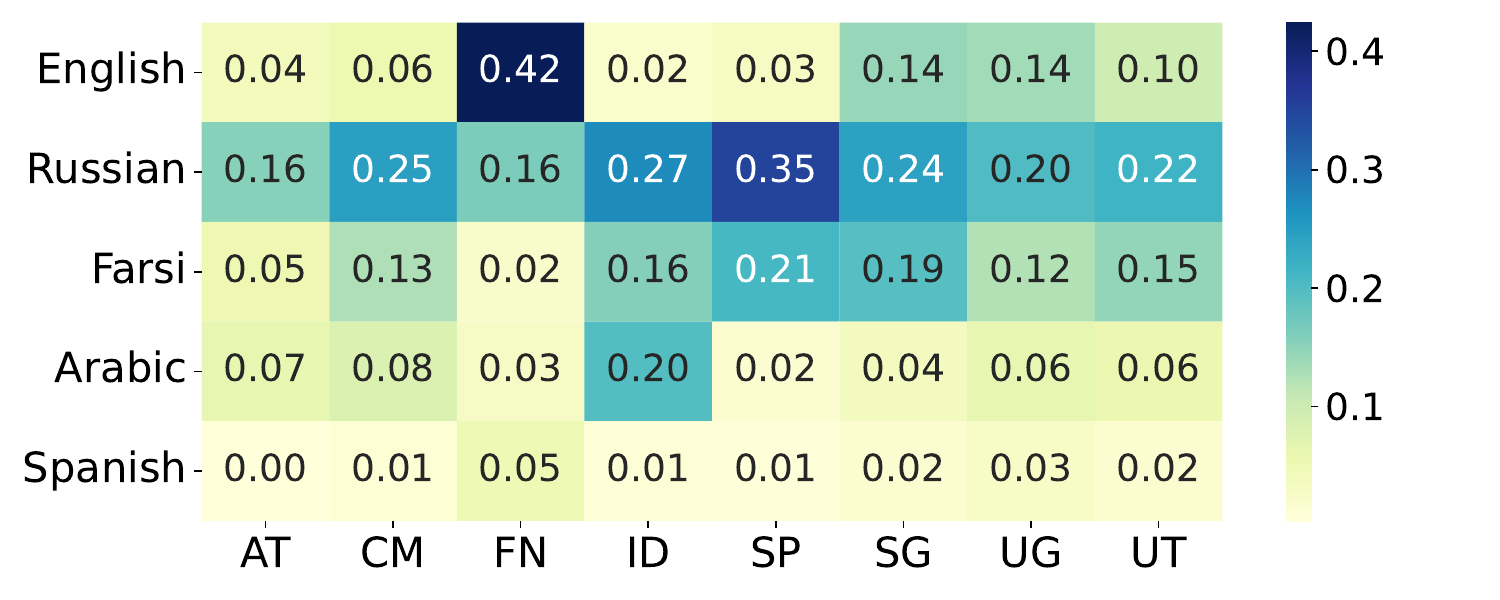}
    \caption{Distribution of bot domains and languages. Showing only the top five languages, so the columns do not add up to 1.}
    \label{fig:domain_language}
\end{figure}

We next compare the language (in messages) against the domains we discover in \S\ref{subsubsec:domains_results}. 
Figure~\ref{fig:domain_language}
shows the distribution of bot domains (x-axis) and the top five languages (y-axis). We find
that the primary language significantly differs by domain. 
English is only dominant in Finance (FN), 42\%, compared to 2\% in Ideology (ID) while Russian is widely used in many domains.
We also observe a relatively high concentration of Arabic in Ideology and Farsi in Shopping (SP) and
Social \& Gaming (SG). 
Underground (UG) has a diverse set of languages.

\subsection{Duration and Reuse}
\label{subsec:duration}
\begin{figure}
    \centering
    \includegraphics[width=1\linewidth]{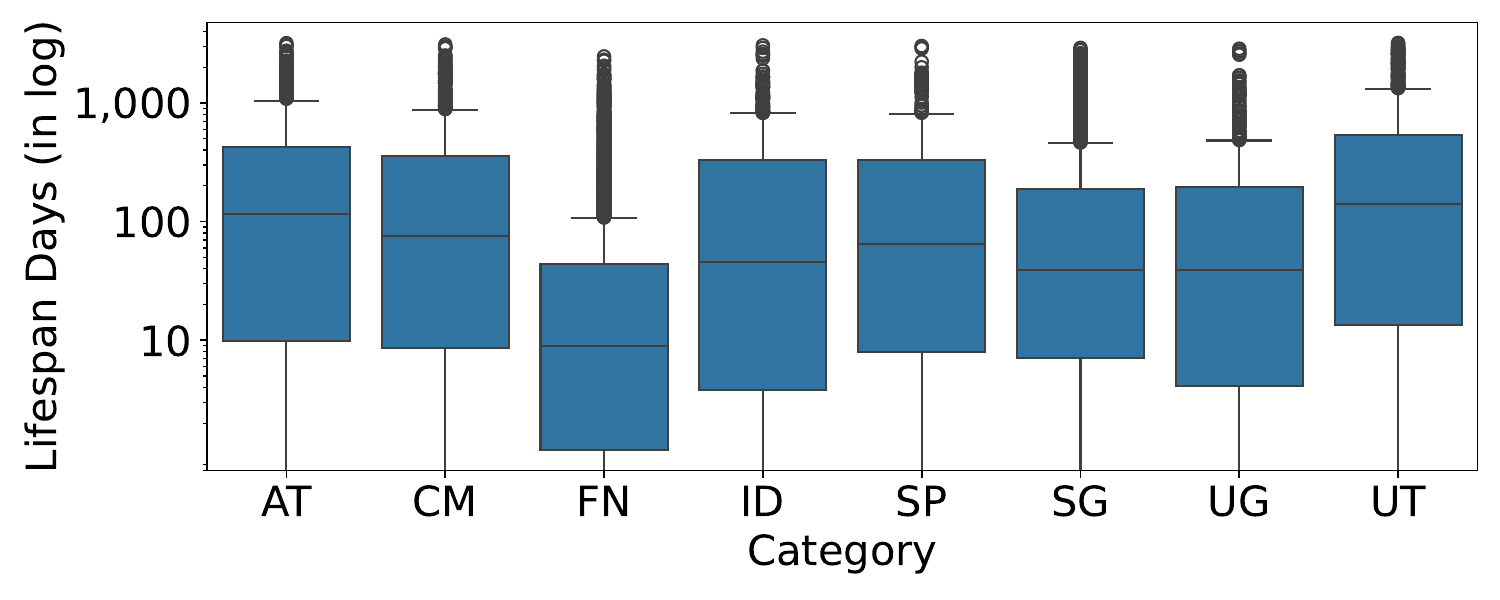}
    \caption{Distribution of bot duration per category.}
    \label{fig:longevity_per_category}
\end{figure}

We next analyze how long the communities use bots and whether they reuse them.
We also look at the aggregated trend of bot usage over time. We define bot
duration as the time difference between the first and last time the bot is
mentioned in messages, which reflects the time period during which the bot is
actively used. We exclude bots that have been mentioned only once.
The average (median) duration is 178 days (21 days).
Figure~\ref{fig:longevity_per_category} shows the distribution of bot duration
per category. Expectedly, Utility (UT) and Admin Tools (AT) have long durations (median
140, 115 days, respectively) as those bots provide relatively static services.
On the other hand, Finance has the shortest duration (median 9 days),
likely due to 1) the short life cycle of investment opportunities (e.g.,
airdrop), 2) the malicious use, as evidenced by a high number of scam
warnings in \S\ref{subsec:malicious_use}, and 3) the reuse of the same bot with
different usernames. 

We indeed observe that some bot developers appear to recreate the bot with the different username (i.e., reuse). 
About 3.4\% (1,074 bots, including the original bot) have the exact same descriptions (with the text length of more than 10 characters).
Of those, 43\% belong to the Finance category. 
At maximum, one bot has been reused 30 times. 
We also look at the similarity of usernames that have the same description.
Bot developers either 1) use the same username with a slight modification with numbers and characters (e.g., \textit{aidropbot}, \textit{aidrop2bot}, ...), 2) iterate different keywords (e.g., swapping financial asset names), or 3) use completely different ones. 
To quantify the similarity of usernames, we calculate the Levenshtein distance\footnote{Number of operations needed to replace one string with another, including swapping.} for every pair in the group that shares the same description.
385 (35.9\%) of bots have a username that is similar to at least one username in the group (with a distance of 1 or 2), which is likely to fall into the first category.

\begin{figure}
    \centering
    \includegraphics[width=1\linewidth]{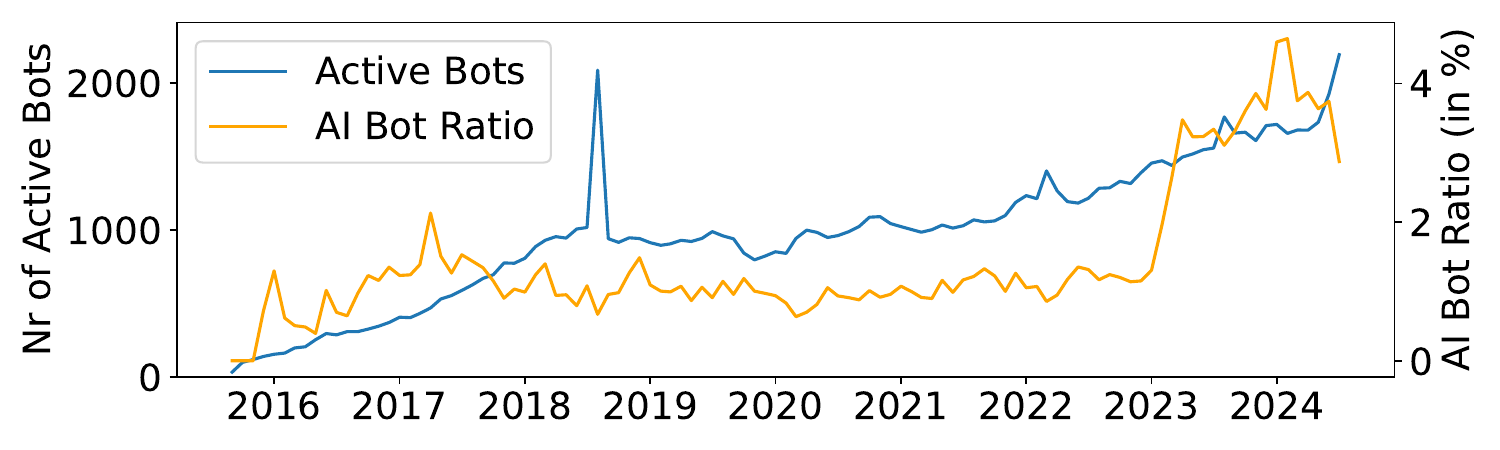}
    \caption{Number of monthly active bots over time with the ratio of AI-related bots.}
    \label{fig:nr_active_bots}
\end{figure}

We further look at the overall trend in the number of active bots.
Figure~\ref{fig:nr_active_bots} shows the number of monthly active bots over
time, as well as the ratio (\%) of AI-related bots (as defined in
\S\ref{subsec:functionality}). We define a bot as active if the
bot has been mentioned at least once in that month. Generally, the number of
active bots has been increasing over time in our dataset. The significant spike in
Aug. 2018 is likely caused by one channel that advertised 1,016 bots in three days.
Bots that provide AI-related services have significantly increased since early 2023, which
coincides with the release of GPT-4 in March 2023. 
Telegram's user friendly interface may have contributed to the adoption of AI for mobile-users, 
which we discus in \S\ref{sec:discussion}. 

\subsection{Channel Topology}
\label{subsec:distribution}
We next look at the network topology of the community surrounding each bot. We prepare a
set of \textit{channels} that mention the same bot and investigate if there is any
interaction within those channels (i.e., a channel mentioning another channel through URLs). We construct an undirected graph
for the set of channels that mentions the same bot and draw an edge
if one channel mentions another channel (i.e., no bot in the
graph). To represent the connectivity of the graph, we use two metrics: (1)
\textit{density}, which is the ratio of the number of edges to the number of
possible edges, and (2) \textit{average degree}, which measures the average
number of connections per node in the graph. For both metrics, higher values
indicate a more connected network (i.e., a dense community).

\begin{figure}
    \centering
    \includegraphics[width=1\linewidth]{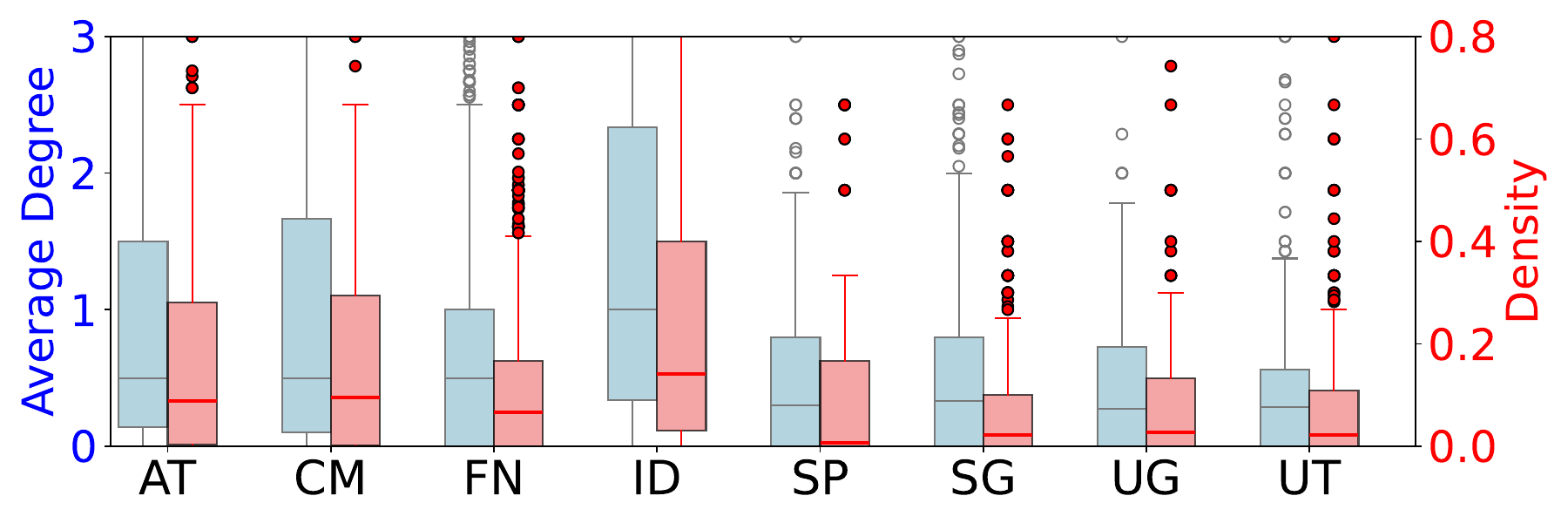}
    \caption{Distribution of channel network metrics per bot domain (blue: average degree, red: density).}
    \label{fig:channel_network}
\end{figure}

We only consider bots that are mentioned by at least 4 channels to exclude isolates, dyads, and triads.
Figure~\ref{fig:channel_network} shows the distribution of density and average
degree per domain. We find that Ideology (ID) bots have the highest density and
average degree, whereas Utility (UT) bots have lower scores in those metrics. This
implies that communities around Ideology bots are more interconnected. Due to
their coordinated activities related to religion, politics, war, and social
movements, their bots are likely designed for a specific community. Utility
bots, on the other hand, are likely mentioned by a diverse set of channels with
less interaction among them, and are mostly designed for individual use.

\begin{figure}
    \centering
    \begin{minipage}{0.3\linewidth}
        \centering
        \includegraphics[width=1\linewidth]{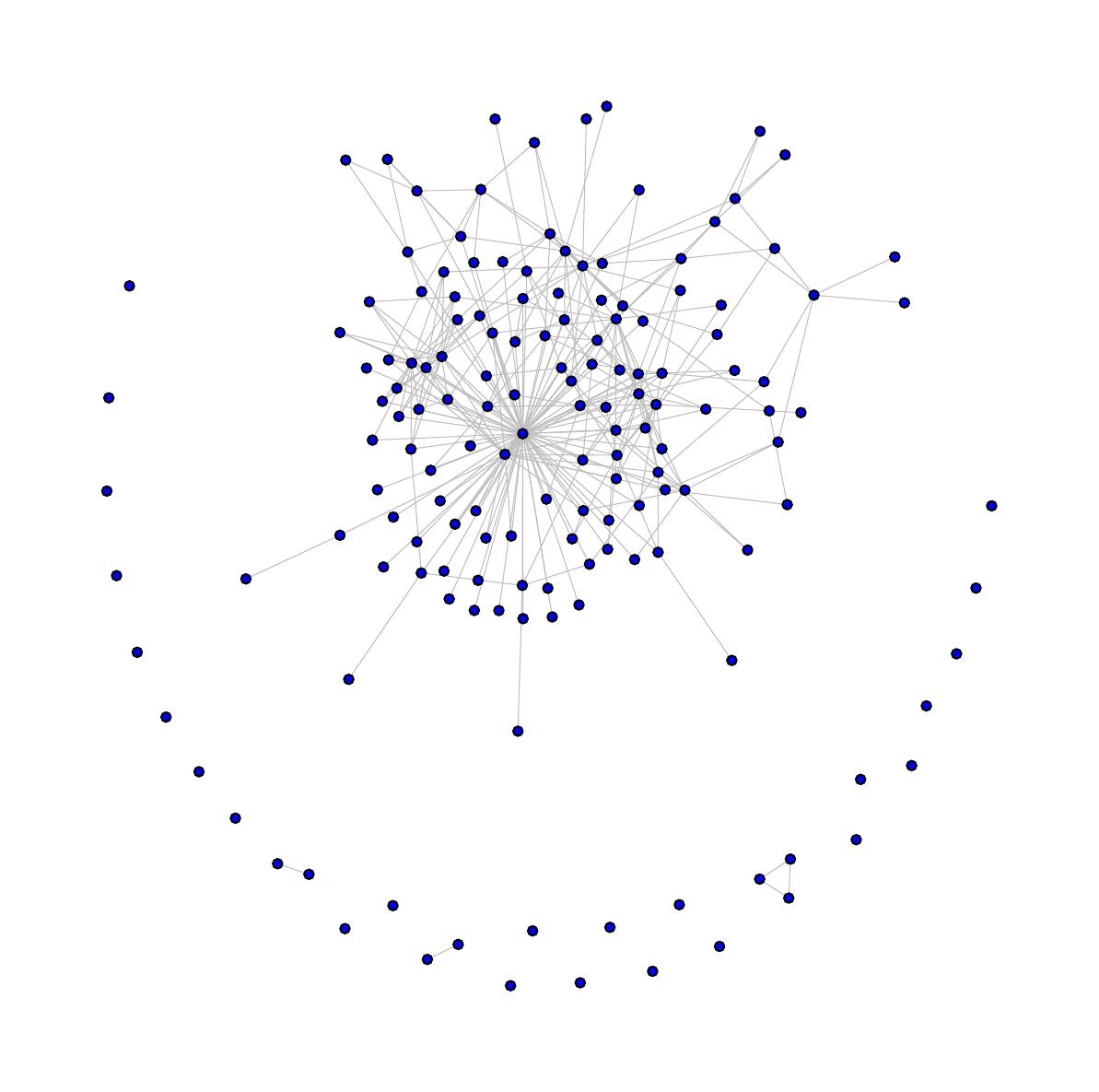}
    \end{minipage}
    \hfill
    \begin{minipage}{0.3\linewidth}
        \centering
        \includegraphics[width=1\linewidth]{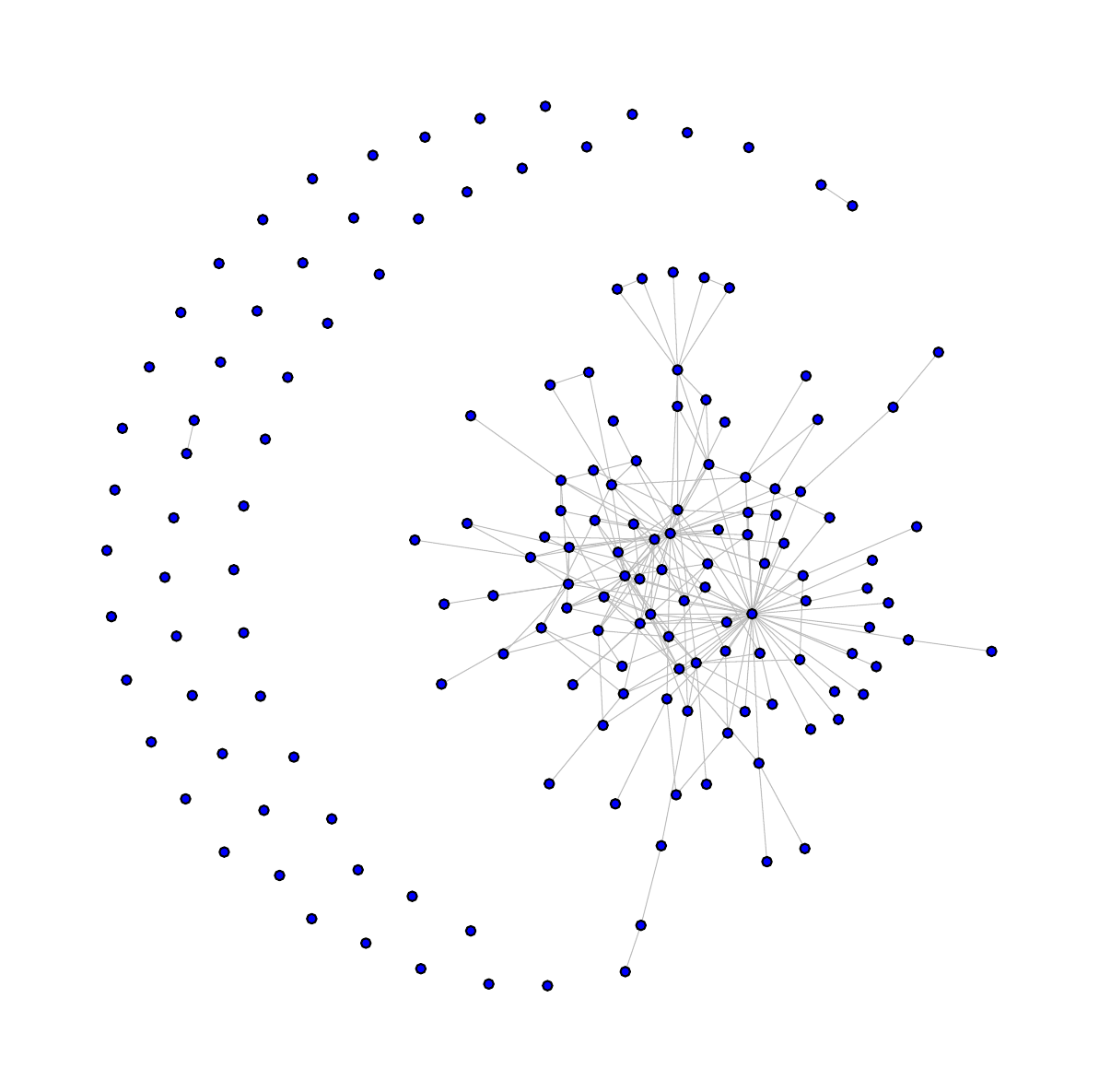}
    \end{minipage}
    \hfill
    \begin{minipage}{0.3\linewidth}
        \centering
        \includegraphics[width=1\linewidth]{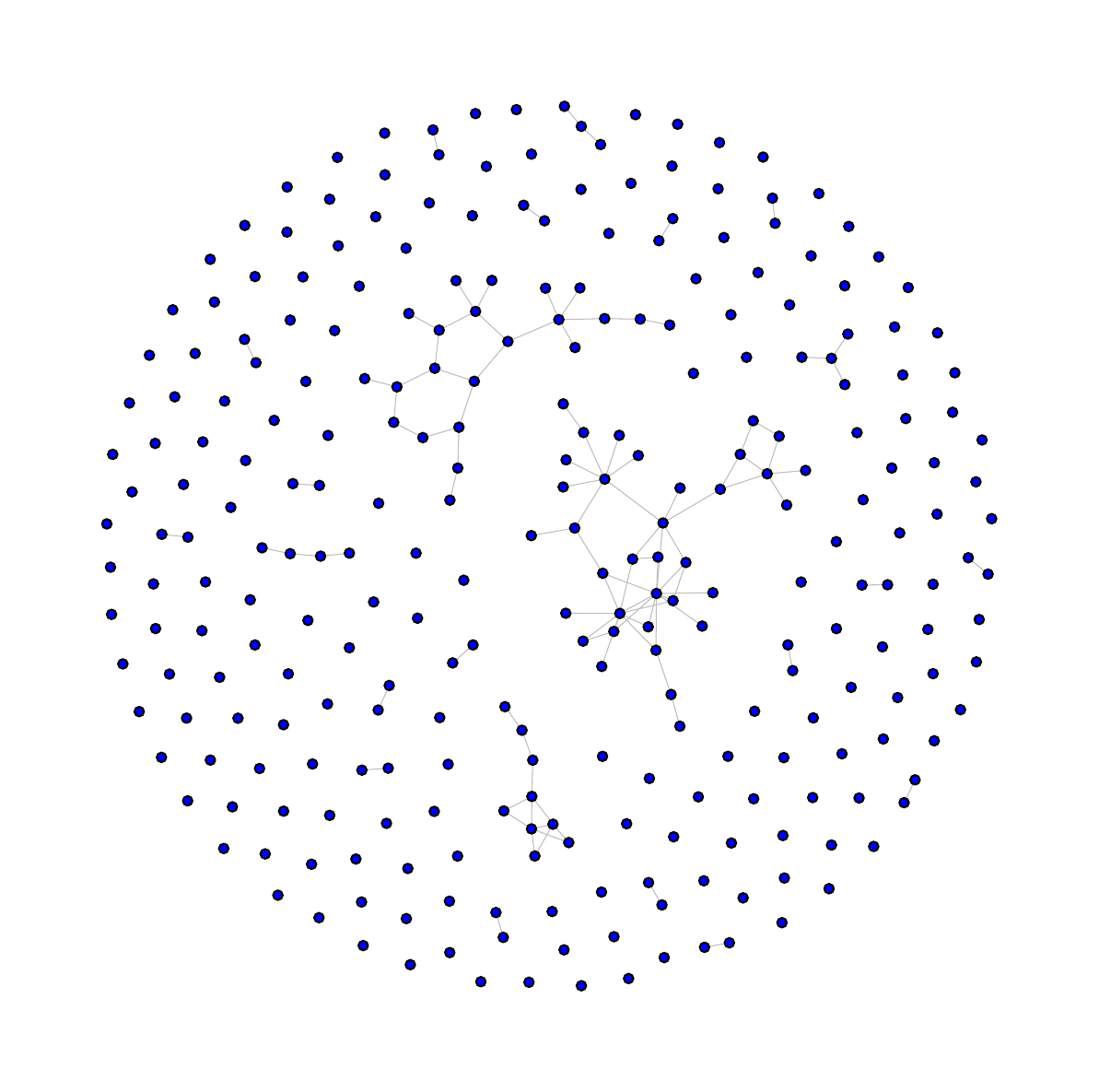}
    \end{minipage}
    \caption{Channel networks for three example bots.}
    \label{fig:example_channel_networks}
\end{figure}

Figure~\ref{fig:example_channel_networks} shows the three example channel networks; the left and the middle figures are in Ideology and the right is in Utility.
We keep the isolates in those figures.
The left Ideology bot appears to collect information on Russia's military
operations to Ukraine, and the middle Ideology bot seems to distribute an
Iranian leader's publications. All the well-connected channels appear to be
related to those specific topics. The right Utility bot helps users discover other
bots. While some Brazilian communities happen to use the same bot, which creates
a small cluster, the rest is highly disconnected. Bots in
different domains exhibit different purposes and usage patterns.

\section{Discussion}
\label{sec:discussion}
This section highlights how our findings can inform future research, platform policy, and regulation.
We also discuss the limitations of our study.

\subsection{Implications}
\label{subsec:implication}
We provide the first characterization of Telegram bots use at large scale. While
the majority of bots are benign, we uncovered bots which are used for illegal or concerning purposes, such as
carrying out fraud (n=1,331, 4\%), commercializing illicit goods and services, or providing
access to questionable services, such as AI non-consensual deepfakes (n=1,539, 5\%). Unlike
traditional darknet marketplaces or forums on the Tor network that require
desktop environments, Telegram offers a mobile-first, easier-to-use (for both
developers and end users) platform that lowers the barrier to provide
questionable or outright illicit services. We theorize that Telegram fills a gap
between clearnet websites and hidden services. The growth of ``gray'' AI services
and the abundance of cryptocurrency/blockchain-related bots support this hypothesis.

Our findings also confirm that bots are a crucial part of the infrastructure
that process payments, expand channels, and provide access to content - both
benign and malicious. 
We posit that future studies on Telegram should take into account the role of
bots in supporting these activities; we provide a comprehensive dataset for
researchers to do so. We also suggest moderators and law enforcement consider
bots as a potential intervention point to disable monetization and expansion.
Focusing on bots could be more effective than targeting users or channels simply
because the number of bots is significantly smaller than the number of channels
or users.

We offer several suggestions to moderate bots. First, our method of collecting
the bot's command list and the messages mentioning bots is effective in profiling 
bot operations (\S\ref{subsec:bot_interaction}). Not only is it important to uncover
better information, but also 36\% of existing bots do not have a description, so
it would not be possible to infer their use without interaction. Second, we
observed an association between languages and domains (\S\ref{subsec:languages}). Language may indicate
geographic location, which could help focus law enforcements' efforts. Third,
some bots (e.g., such as those in Finance) are short-lived, and often reused,
which makes moderation challenging (\S\ref{subsec:duration}). Telegram could reduce the default number of
bots users can create (currently 20) or freeze other bots created by the same
user if one of the bots is flagged (i.e., guilt-by-association). Clustering bots
with similar description and usernames could also help uncover more bots.
Fourth, purpose-specific bots (as opposed to generic utility ones) tend to serve
closely knit communities, which could be another avenue for moderation (\S\ref{subsec:distribution}). Finally, bots
offering payment and referral functionalities should be treated with caution,
especially in the Finance domain (\S\ref{subsec:malicious_use}).

\subsection{Limitations}
\label{subsec:limitation_futurework}
Our dataset may not constitute a representative sample of Telegram. Our dataset
does not contain any private channels that require invitations, which likely
causes us to underestimate Underground domains. Our bot sample, extracted from channels, 
groups, and third-party lists, may also be biased towards popular bots.
Lastly, we only send universal commands to bots and do not have any multi-stage interaction. 
For instance, we could not identify the types
of payment methods or AI models at scale because some bots only disclose them
after a few interactions. 
Beyond the basic commands, Telegram bots accept a wide range of input methods, from inline queries (triggered from anywhere in Telegram) to buttons, plain text, and non-textual inputs such as images or location data. 
Multi-stage interaction also requires to mange the conversation state/flow, which will significantly increase the data collection costs and potentially limit the number of bots we could interact with. 
This study prioritized a broader sample over in-depth exploration. 
Future work could potentially use LLMs to have a longer conversation with bots while adhering to any ethical concerns discussed below. 
\section{Conclusion}
\label{sec:conclusion}
This paper is the first to look at the programmable aspects of Telegram, namely \textit{bots}. 
We develop a novel system to continuously collect Telegram messages and interact with bots.
This collected data enables us to analyze their functionalities and usage patterns at scale.
Despite many legitimate use cases (e.g., crowdsourcing), some bots help illicit activities such as financial scams, cybercrime, and non-consensual image sharing.
This paper alerts various stakeholders, including researchers, Telegram, and law enforcement, to recognize bots as an emerging software infrastructure. %
\section*{Ethical Statement}
\label{subsec:ethics}
We only collected publicly accessible data on Telegram using their official
APIs, complying with Telegram's ToS. The Telegram API allows reading channel public messages without joining
them, minimizing impact on the community. We did not collect any private channels and did not attempt to deanonymize users. Under such circumstances, informed consent may be waived~\cite{britishethics}.  

To avoid downloading malicious or sensitive information, we did not collect any
media files. However, our dataset contains some problematic channels, such as
financial scams or underground services. Given the size of the dataset (up to
700 GB), we cannot eliminate the possibility that some users publicly disclose
sensitive information (e.g., personally identifiable information). Malicious actors could also misuse our dataset to find bots that facilitate illicit activities. 
In this way, our data was released, guided by \citet{fair} and \citet{gebru2021datasheets}, under CC Attribution No Derivatives 4.0 International License, but made only available to researchers who agree to our data usage
policy through Zenodo\footnote{The request for the dataset can be made at \url{https://zenodo.org/records/17281308} after logging into your Zenodo account} to minimize such harms.

While most of our data collection is passive (e.g., reading messages from
channels), we interact with bots. We consider bots as public
API endpoints and only use Telegram's official API to send commands. We
limit ourselves to sending two basic commands (e.g., \texttt{/start}, \texttt{/help}), which, we believe, falls 
within acceptable bot use.
The IRB (i.e., ethics committee) at our institution only reviews human subject studies, which does not apply to our case.
\section*{Acknowledgement}
We confirm that all text in this paper was written by the authors. 
AI-based writing assistants (e.g., Grammarly, Claude) were used solely for grammar and spelling checks and to improve the clarity of the author-written text. 
We sometimes used Copilot for automatic code completion (primarily for figures) to reduce typing errors. 
However, we manually verified all lines of codes. 

This work was supported by the Cylab Presidential Fellowship, CyLab Security and Privacy Institute seed funding grant, the Nakajima Foundation, King’s College Cambridge, and the Cambridge Trust, European Research Council (ERC) under the European Union’s Horizon 2020 research and innovation programme, grant No 949127.  
\bibliography{aaai2026}

@inproceedings{tgdataset,
  author = {La Morgia, Massimo and Mei, Alessandro and Mongardini, Alberto Maria},
  title = {TGDataset: Collecting and exploring the largest Telegram channels dataset},
  year = {2025},
  publisher = {Association for Computing Machinery},
  address = {New York, NY, USA},
  url = {https://doi.org/10.1145/3690624.3709397},
  doi = {10.1145/3690624.3709397},
  booktitle = {Proceedings of the 31st ACM SIGKDD Conference on Knowledge Discovery and Data Mining V.1},
  pages = {2325–2334},
  numpages = {10},
  location = {Toronto ON, Canada},
  series = {KDD'25}
}

@inproceedings{pushshift,
  title={The Pushshift Telegram dataset},
  author={Baumgartner, Jason and Zannettou, Savvas and Squire, Megan and Blackburn, Jeremy},
  booktitle={Proceedings of the international AAAI conference on web and social media},
  volume={14},
  pages={840--847},
  year={2020}
}

@inproceedings{kireev2025telegram,
  title={A Telegram dataset of propaganda and its moderation},
  author={Kireev, Klim and Mykhno, Yevhen and Troncoso, Carmela and Overdorf, Rebekah},
  booktitle={Proceedings of the International AAAI Conference on Web and Social Media},
  volume={19},
  pages={2510--2518},
  year={2025}
}

@article{guo2024beyond,
  author = {Guo, Yanhui and Wang, Dong and Wang, Liu and Fang, Yongsheng and Wang, Chao and Yang, Minghui and Liu, Tianming and Wang, Haoyu},
  title = {Beyond app markets: Demystifying underground mobile app distribution via Telegram},
  year = {2024},
  issue_date = {December 2024},
  publisher = {Association for Computing Machinery},
  address = {New York, NY, USA},
  volume = {8},
  number = {3},
  url = {https://doi.org/10.1145/3700432},
  doi = {10.1145/3700432},
  journal = {Proceedings of the ACM on Measurement and Analysis of Computing Systems},
  month = {dec},
  articleno = {33},
  numpages = {25},
}

@inproceedings{blas2025unearthing,
  title={Unearthing a billion Telegram posts about the 2024 US presidential election: Development of a public dataset},
  author={Blas, Leonardo and Luceri, Luca and Ferrara, Emilio},
  year = {2025},
  isbn = {9798400713316},
  publisher = {Association for Computing Machinery},
  address = {New York, NY, USA},
  url = {https://doi.org/10.1145/3701716.3715297},
  doi = {10.1145/3701716.3715297},
  booktitle = {Companion Proceedings of the ACM on Web Conference 2025},
  pages = {729–732},
  numpages = {4},
  location = {Sydney NSW, Australia},
  series = {WWW '25}
}

@inproceedings{gangopadhyay2025telescope,
  title={TeleScope A longitudinal dataset for investigating online discourse and information interaction on Telegram},
  author={Gangopadhyay, Susmita and Dessi, Danilo and Dimitrov, Dimitar and Dietze, Stefan},
  booktitle={Proceedings of the International AAAI Conference on Web and Social Media},
  volume={19},
  pages={2423--2433},
  year={2025}
}

@article{bawa2025telegram, 
  title={Telegram as a Battlefield: Kremlin-Related Communications During the Russia-Ukraine Conflict}, 
  volume={19}, 
  url={https://ojs.aaai.org/index.php/ICWSM/article/view/35939}, 
  DOI={10.1609/icwsm.v19i1.35939}, 
  journal={Proceedings of the International AAAI Conference on Web and Social Media}, 
  author={Bawa, Apaar and Kursuncu, Ugur and Achilov, Dilshod and Shalin, Valerie L. and Agarwal, Nitin and Akbas, Esra}, 
  year={2025}, 
  pages={2361-2370}
}

@article{nizzoli2020charting,
  title={Charting the landscape of online cryptocurrency manipulation},
  author={Nizzoli, Leonardo and Tardelli, Serena and Avvenuti, Marco and Cresci, Stefano and Tesconi, Maurizio and Ferrara, Emilio},
  journal={IEEE Access},
  volume={8},
  pages={113230--113245},
  year={2020},
  publisher={IEEE}
}

@inproceedings{xu2019anatomy,
  title={The anatomy of a cryptocurrency Pump-and-Dump scheme},
  author={Xu, Jiahua and Livshits, Benjamin},
  booktitle={28th USENIX Security Symposium (USENIX Security 19)},
  pages={1609--1625},
  year={2019}
}

@article{mirtaheri2021identifying,
  title={Identifying and analyzing cryptocurrency manipulations in social media},
  author={Mirtaheri, Mehrnoosh and Abu-El-Haija, Sami and Morstatter, Fred and Ver Steeg, Greg and Galstyan, Aram},
  journal={IEEE Transactions on Computational Social Systems},
  volume={8},
  number={3},
  pages={607--617},
  year={2021},
  publisher={IEEE}
}

@inproceedings{marjanov2026stayin,
  author = {Marjanov, Tina and Tsuchiya, Taro and Ioannidis, Konstantinos and Hughes, Jack and Christin, Nicolas and Hutchings, Alice},
  booktitle={Proceedings of the 35th USENIX Security Symposium (USENIX Security'26)}, 
  title = {Stayin’ Alive: How Global Stolen Data Markets Thrive on Telegram},
  year = {2026},
}

@article{roy2024darkgram,
  title={DarkGram: A large-scale analysis of cybercriminal activity channels on Telegram},
  author={Roy, Sayak Saha and Vafa, Elham Pourabbas and Khanmohamaddi, Kobra and Nilizadeh, Shirin},
  booktitle={Proceedings of the 34th USENIX Security Symposium (USENIX Security'25)},
  year={2025}
}

@article{gao2020tracking,
  title={Tracking counterfeit cryptocurrency end-to-end},
  author={Gao, Bingyu and Wang, Haoyu and Xia, Pengcheng and Wu, Siwei and Zhou, Yajin and Luo, Xiapu and Tyson, Gareth},
  journal={Proceedings of the ACM on Measurement and Analysis of Computing Systems},
  volume={4},
  number={3},
  pages={1--28},
  year={2020},
  publisher = {Association for Computing Machinery},
  address = {New York, NY, USA},
  url = {https://doi.org/10.1145/3428335},
  doi = {10.1145/3428335},
  articleno = {50},
}

@inproceedings{cernera2023token,
  title={Token spammers, rug pulls, and sniper bots: An analysis of the ecosystem of tokens in Ethereum and in the Binance Smart Chain (BNB)},
  author={Cernera, Federico and La Morgia, Massimo and Mei, Alessandro and Sassi, Francesco},
  booktitle={32nd USENIX Security Symposium (USENIX Security 23)},
  pages={3349--3366},
  year={2023}
}

@inproceedings{bijmans2021catching,
  title={Catching phishers by their bait: Investigating the Dutch phishing landscape through phishing kit detection},
  author={Bijmans, Hugo and Booij, Tim and Schwedersky, Anneke and Nedgabat, Aria and van Wegberg, Rolf},
  booktitle={30th USENIX Security Symposium (USENIX Security 21)},
  pages={3757--3774},
  year={2021}
}

@misc{nigam2018user, 
    title={User Telegram bot to remotely control infected devices by malware}, 
    howpublished={\url{https://unit42.paloaltonetworks.com/unit42-telerat-another-android-trojan-leveraging-telegrams-bot-api-to-target-iranian-users/}}, 
    journal={}, 
    publisher={}, 
    author={Nigam, Ruchna and Wilhoit, Kyle}, 
    year={2018}, 
    month={},
    note = {Accessed Sep. 23rd, 2024}
}

@misc{buyukkaya2024onnx, 
    title={ONNX Store: Phishing-as-a-service platform targeting financial Institution}, 
    howpublished={\url{https://blog.eclecticiq.com/onnx-store-targeting-financial-institution}}, 
    journal={}, 
    publisher={}, 
    author={Büyükkaya, Arda}, 
    year={2024}, 
    month={},
    note = {Accessed Sep. 23rd, 2024}
}

@inproceedings{ricaldi2025uncovering,
    title={Uncovering the Trust Signals Supporting Telegram’s Cybercrime Economy},
    author={Ricaldi, Roy and Marjanov, Tina and Allodi, Luca and Hutchings, Alice},
    booktitle={2025 APWG Symposium on Electronic Crime Research (eCrime)},
    pages={1--17},
    year={2025},
    organization={IEEE}
}

@inproceedings{marjanov2025sok,
  title={SoK: Digging into the digital underworld of stolen data markets},
  author={Marjanov, Tina and Hutchings, Alice},
  booktitle={2025 IEEE Symposium on Security and Privacy (SP)},
  pages={1--18},
  year={2025},
  organization={IEEE}
}

@inproceedings{han2025characterizing,
  title={Characterizing the MrDeepFakes sexual deepfake marketplace},
  author={Han, Catherine and Li, Anne and Kumar, Deepak and Durumeric, Zakir},
  booktitle={34th USENIX Security Symposium (USENIX Security 25)},
  year={2025}
}

@inproceedings{gibson2025analyzing,
  title={Analyzing the AI nudification application ecosystem},
  author={Gibson, Cassidy and Olszewski, Daniel and Brigham, Natalie Grace and Crowder, Anna and Butler, Kevin RB and Traynor, Patrick and Redmiles, Elissa M and Kohno, Tadayoshi},
  booktitle={34th USENIX Security Symposium (USENIX Security 25)},
  year={2025}
}

@article{semenzin2020use,
  title={The use of Telegram for non-consensual dissemination of intimate images: Gendered affordances and the construction of masculinities},
  author={Semenzin, Silvia and Bainotti, Lucia},
  journal={Social Media+ Society},
  volume={6},
  number={4},
  year={2020},
}

@article{kireev2025characterizing,
  title={Characterizing and detecting propaganda-spreading accounts on Telegram},
  author={Kireev, Klim and Mykhno, Yevhen and Troncoso, Carmela and Overdorf, Rebekah},
  journal={34th USENIX Security Symposium (USENIX Security 25)},
  year={2025}
}

@article{imperati2025conspiracy,
  title={The conspiracy money machine: Uncovering Telegram’s conspiracy channels and their profit model},
  author={Imperati, Vincenzo and La Morgia, Massimo and Mei, Alessandro and Mongardini, Alberto Maria and Sassi, Francesco},
  year={2025},
  journal={34th USENIX Security Symposium (USENIX Security 25)},
}

@inproceedings{nikkhah2018telegram,
  title={Telegram as an immigration management tool},
  author={Nikkhah, Sarah and Miller, Andrew D and Young, Alyson L},
  booktitle={Companion of the 2018 ACM Conference on Computer Supported Cooperative Work and Social Computing},
  pages={345--348},
  year={2018}
}

@inproceedings{hanley2024partial,
  title={Partial mobilization: Tracking multilingual information flows amongst Russian media outlets and Telegram},
  author={Hanley, Hans WA and Durumeric, Zakir},
  booktitle={Proceedings of the International AAAI Conference on Web and Social Media},
  pages={528--541},
  year={2024}
}

@article{alrhmoun2024automating,
  title={Automating terror: The role and impact of telegram bots in the Islamic State’s online ecosystem},
  author={Alrhmoun, Abdullah and Winter, Charlie and Kert{\'e}sz, J{\'a}nos},
  journal={Terrorism and Political Violence},
  volume={36},
  number={4},
  pages={409--424},
  year={2023},
  publisher={Taylor \& Francis}
}

@article{ng2024exploratory,
  title={An exploratory analysis of COVID bot vs human disinformation dissemination stemming from the Disinformation Dozen on Telegram},
  author={Ng, Lynnette Hui Xian and Kloo, Ian and Clark, Samantha and Carley, Kathleen M},
  journal={Journal of Computational Social Science},
  volume={7},
  pages={695--720},
  year={2024},
  publisher={Springer}
}

@inproceedings{de2016chatting,
  title={Chatting with Arduino platform through Telegram bot},
  author={De Oliveira, Juan Carlos and Santos, Danilo Henrique and Neto, M{\'a}rio Popolin},
  booktitle={2016 IEEE International Symposium on Consumer Electronics},
  year={2016},
  organization={IEEE}
}

@inproceedings{franco2024characterizing,
  title={Characterizing non-consensual intimate image abuse on Telegram groups and channels},
  author={Franco, Mirko and Gaggi, Ombretta and Palazzi, Claudio E},
  booktitle={Proceedings of the 4th International Workshop on Open Challenges in Online Social Networks},
  pages={26--32},
  year={2024}
}

@inproceedings{steffen2025more,
  title={More than Memes: A Multimodal Topic Modeling Approach to Conspiracy Theories on Telegram},
  author={Steffen, Elisabeth},
  booktitle={Proceedings of the International AAAI Conference on Web and Social Media},
  volume={19},
  pages={1831--1844},
  year={2025}
}

@inproceedings{perlo2025topic,
  title={Topic-wise exploration of the Telegram group-verse},
  author={Perlo, Alessandro and Paoletti, Giordano and Jha, Nikhil and Vassio, Luca and Almeida, Jussara and Mellia, Marco},
  booktitle={Companion Proceedings of the ACM on Web Conference 2025},
  pages={1792--1801},
  year={2025}
}

@article{cohen1960coefficient,
  title={A coefficient of agreement for nominal scales},
  author={Cohen, Jacob},
  journal={Educational and Psychological Measurement},
  volume={20},
  pages={37--46},
  year={1960},
}

@misc{danilk2025langdetect,
  title={langdetect 1.0.9},
  author={Danilk, Michal},
  year={2025},
  howpublished={\url{https://pypi.org/project/langdetect/}},
  note={Accessed Oct. 7th, 2025}
}

@misc{britishethics,
  title = {Statement of ethics},
  author = {{British Society of Criminology}},
  year = {2015},
  howpublished = {\url{https://www.britsoccrim.org/ethics/}}
}

@misc{fair,
    title="The FAIR data principles",
    year = {2020},
    author= {FORCE11},
    howpublished={\url{https://force11.org/info/the-fair-data-principles/}}
}

@article{gebru2021datasheets,
  title={Datasheets for datasets},
  author={Gebru, Timnit and Morgenstern, Jamie and Vecchione, Briana and Vaughan, Jennifer Wortman and Wallach, Hanna and Iii, Hal Daum{\'e} and Crawford, Kate},
  journal={Communications of the ACM},
  volume={64},
  number={12},
  pages={86--92},
  year={2021},
  publisher = {Association for Computing Machinery},
  address = {New York, NY, USA},
}

@misc{lieber2023stolen, 
    title={Stolen checks are for sale online. We called some of the victims.}, 
    howpublished={\url{https://www.nytimes.com/2023/12/09/business/stolen-checks-telegram.html}}, 
    journal={}, 
    publisher={New York Times}, 
    author={Lieber, Ron}, 
    year={2023}, 
    month={},
    note = {Accessed Sep. 24th, 2025}
}

@misc{gebrekidan2023thescammer, 
    title={The scammer’s manual: how to launder money and get away with it.}, 
    howpublished={\url{https://www.nytimes.com/2025/03/23/world/asia/cambodia-money-laundering-huione.html}}, 
    journal={}, 
    publisher={New York Times}, 
    author={Gebrekidan, Selam and Dong, Joy}, 
    year={2025}, 
    month={},
    note = {Accessed Sep. 24th, 2025}
}

@misc{burgess2020telegram, 
    title={Telegram still hasn’t removed an AI bot that’s abusing women}, 
    howpublished={\url{https://www.wired.com/story/telegram-still-hasnt-removed-an-ai-bot-thats-abusing-women/}}, 
    journal={}, 
    publisher={WIRED}, 
    author={Burgess, Matt}, 
    year={2020}, 
    month={},
    note = {Accessed Sep. 24th, 2025}
}

@misc{mozur2024telegram, 
    title={How Telegram became a playground for criminals, extremists and terrorists}, 
    howpublished={\url{https://www.nytimes.com/2024/09/07/technology/telegram-crime-terrorism.html}}, 
    journal={}, 
    publisher={New York Times}, 
    author={Mozur, Paul and Satariano, Adam and Krolik, Aaron and Myers, Steven}, 
    year={2024}, 
    month={},
    note = {Accessed Sep. 24th, 2025}
}

@misc{telegram2025faq, 
    title={Telegram FAQ}, 
    author={Telegram}, 
    year={2025},
    howpublished={\url{https://telegram.org/faq/}}, 
    note = {Accessed Oct. 4th, 2025}
}

@misc{telegram2025widget,
  title={Telegram login widget},
  author={Telegram},
  year={2025},
  howpublished={\url{https://core.telegram.org/widgets/login}},
  note={Accessed Sep. 19th, 2025}
}

@misc{telegram2025passport,
  title={Telegram passport blog},
  author={Telegram},
  year={2025},
  howpublished={\url{https://telegram.org/blog/passport}},
  note={Accessed Sep. 19th, 2025}
}

@misc{telegram2025browser,
  title={Telegram browser, mini app store, gifting stars and more},
  author={Telegram},
  year={2025},
  howpublished={\url{https://telegram.org/blog/w3-browser-mini-app-store}},
  note={Accessed Sep. 19th, 2025}
}

@misc{telegram2025star,
  title={Telegram stars: pay for digital goods and more},
  author={Telegram},
  year={2025},
  howpublished={\url{https://telegram.org/blog/telegram-stars}},
  note={Accessed Sep. 19th, 2025}
}

@misc{telegram2025ton,
  title={TON: the open network},
  author={Telegram},
  year={2025},
  howpublished={\url{https://ton.org/}},
  note={Accessed Sep. 19th, 2025}
}

@misc{telegram2025features,
  title={Telegram bot features},
  author={Telegram},
  year={2025},
  howpublished={\url{https://core.telegram.org/bots/features}},
  note={Accessed Sep. 19th, 2025}
}

\section*{Paper Checklist}

\begin{enumerate}

\item For most authors...
\begin{enumerate}
    \item  Would answering this research question advance science without violating social contracts, such as violating privacy norms, perpetuating unfair profiling, exacerbating the socio-economic divide, or implying disrespect to societies or cultures?
    \answerYes{Yes, we never deanonymize users in our Telegram dataset.}
  \item Do your main claims in the abstract and introduction accurately reflect the paper's contributions and scope?
    \answerYes{Yes.}
   \item Do you clarify how the proposed methodological approach is appropriate for the claims made? 
    \answerYes{Yes, in \S\ref{subsec:dataset_comparison} and \S\ref{sec:methodology}.}
   \item Do you clarify what are possible artifacts in the data used, given population-specific distributions?
    \answerYes{Yes, our dataset and analysis code.}
  \item Did you describe the limitations of your work?
    \answerYes{Yes, in \S\ref{subsec:limitation_futurework}.}
  \item Did you discuss any potential negative societal impacts of your work?
    \answerYes{Yes, in Ethical Statement.}
      \item Did you discuss any potential misuse of your work?
    \answerYes{Yes, in Ethical Statement.}
    \item Did you describe steps taken to prevent or mitigate potential negative outcomes of the research, such as data and model documentation, data anonymization, responsible release, access control, and the reproducibility of findings?
    \answerYes{Yes, in Ethical Statement.}
  \item Have you read the ethics review guidelines and ensured that your paper conforms to them?
    \answerYes{Yes.}
\end{enumerate}

\item Additionally, if your study involves hypotheses testing...
\begin{enumerate}
  \item Did you clearly state the assumptions underlying all theoretical results?
    \answerNA{NA}
  \item Have you provided justifications for all theoretical results?
    \answerNA{NA}
  \item Did you discuss competing hypotheses or theories that might challenge or complement your theoretical results?
    \answerNA{NA}
  \item Have you considered alternative mechanisms or explanations that might account for the same outcomes observed in your study?
    \answerNA{NA}
  \item Did you address potential biases or limitations in your theoretical framework?
    \answerNA{NA}
  \item Have you related your theoretical results to the existing literature in social science?
    \answerNA{NA}
  \item Did you discuss the implications of your theoretical results for policy, practice, or further research in the social science domain?
    \answerNA{NA}
\end{enumerate}

\item Additionally, if you are including theoretical proofs...
\begin{enumerate}
  \item Did you state the full set of assumptions of all theoretical results?
    \answerNA{NA}
	\item Did you include complete proofs of all theoretical results?
    \answerNA{NA}
\end{enumerate}

\item Additionally, if you ran machine learning experiments...
\begin{enumerate}
  \item Did you include the code, data, and instructions needed to reproduce the main experimental results (either in the supplemental material or as a URL)?
    \answerNA{NA}
  \item Did you specify all the training details (e.g., data splits, hyperparameters, how they were chosen)?
    \answerNA{NA}
     \item Did you report error bars (e.g., with respect to the random seed after running experiments multiple times)?
    \answerNA{NA}
	\item Did you include the total amount of compute and the type of resources used (e.g., type of GPUs, internal cluster, or cloud provider)?
    \answerNA{NA}
     \item Do you justify how the proposed evaluation is sufficient and appropriate to the claims made? 
    \answerNA{NA}
     \item Do you discuss what is ``the cost`` of misclassification and fault (in)tolerance?
    \answerNA{NA}
  
\end{enumerate}

\item Additionally, if you are using existing assets (e.g., code, data, models) or curating/releasing new assets, \textbf{without compromising anonymity}...
\begin{enumerate}
  \item If your work uses existing assets, did you cite the creators?
    \answerYes{Yes, we used two existing datasets (in \S\ref{subsec:collect_channels_messages}).}
  \item Did you mention the license of the assets?
    \answerYes{Yes in Ethical Statement.}
  \item Did you include any new assets in the supplemental material or as a URL?
    \answerYes{Yes, the code and the dataset are released for researchers (following Ethical Statement).}
  \item Did you discuss whether and how consent was obtained from people whose data you're using/curating?
    \answerYes{Yes, in Ethical Statement.}
  \item Did you discuss whether the data you are using/curating contains personally identifiable information or offensive content?
    \answerYes{Yes, in Ethical Statement.}
\item If you are curating or releasing new datasets, did you discuss how you intend to make your datasets FAIR (see \citet{fair})?
\answerYes{Yes, in Ethical Statement.}
\item If you are curating or releasing new datasets, did you create a Datasheet for the Dataset (see \citet{gebru2021datasheets})? 
\answerYes{Yes, we discussed in Ethical Statement.}
\end{enumerate}

\item Additionally, if you used crowdsourcing or conducted research with human subjects, \textbf{without compromising anonymity}...
\begin{enumerate}
  \item Did you include the full text of instructions given to participants and screenshots?
    \answerNA{NA}
  \item Did you describe any potential participant risks, with mentions of Institutional Review Board (IRB) approvals?
    \answerYes{Yes, our work is not considered human subject research.}
  \item Did you include the estimated hourly wage paid to participants and the total amount spent on participant compensation?
    \answerNA{NA}
   \item Did you discuss how data is stored, shared, and deidentified?
   \answerYes{Yes, in Ethical Statement}
\end{enumerate}

\end{enumerate} 
\appendix

\onecolumn
\section{Appendix}
\label{sec:appendix}

\subsection{Bot domain categorization prompt}
\label{sec:appendix_prompt}

\lstdefinestyle{json}{
    basicstyle=\ttfamily, %
    backgroundcolor=\color{gray!5},
    frame=single
}

Choose one category that fits the Telegram bot best from the list below. Do not create a new category.

\begin{description}
    \item[\textbf{Admin Tools}:] Bots that manage groups on behalf of owners, such as membership management, question answering (e.g., using as a point of contact), content moderation, and group statistics.
    
    \item[\textbf{Content \& Media}:] Bots that help distribute or collect educational \& training materials, streaming (music, movies, TV series), and news media, but do not include ones that are from underground markets.
    
    \item[\textbf{Ideology}:] Bots that are used for political campaigns, social movements, or religion-related purposes.
    
    \item[\textbf{Finance}:] Bots that provide access to financial services (including cryptocurrencies and NFTs) such as online wallets, trading, airdrops, mining, or providing financial information, but do not include ones with games or gambling components.
    
    \item[\textbf{Shopping}:] Bots that facilitate online shopping, including selling and buying products, collecting reviews, providing customer service, and product search, but do not include ones from underground markets.
    
    \item[\textbf{Social \& Gaming}:] Bots that facilitate online interactions, including chatting for fun (not Q\&A), dating, games, and gambling.
    
    \item[\textbf{Underground}:] Bots that support underground operations, such as cybercrime (e.g., hacking, stolen data, phishing) and adult content.
    
    \item[\textbf{Utility}:] Bots that provide tools or functions to individual users (not groups), such as using LLM endpoints, developer tools, web search, photo \& video management, health \& fitness management, and QR code generation.
    
    \item[\textbf{Others}:] Bots that do not belong to the categories above or are unknown given the input.
\end{description}

For each bot, we will send you the following:

\begin{lstlisting}[style=json]
{
    "name": <str> the username of the bot which may sometimes 
                  contain important keywords,
    "description": <str> the description of the bot,
    "command_list": <str> the list of commands the bot has,
    "start_response": <str> the response message for the /start command,
    "help_response": <str> the response message for the /help command,
    "message": <str> the messages that include the link to the bot 
    (they may not necessarily be about the bot).
}
\end{lstlisting}

Given an input, I want you to choose one category (\texttt{category}) and produce a one-sentence summary of the bot (\texttt{summary}).

Output format:
\begin{lstlisting}[style=json]
{
    "category": <str>,
    "summary": <str>
}
\end{lstlisting}

\subsection{Bot domain categorization annotation results}
\label{sec:appendix_annotation}

\begin{table}[h]
\centering
\begin{tabular}{@{}ccccc@{}}
\toprule
                 & Annotator 1 & Annotator 2 & GPT-4o & GPT-4o-mini \\ \midrule
Admin Tools      & 0.14         & 0.02         & 0.06   & 0.04        \\
Content \& Media & 0.04         & 0.10         & 0.09   & 0.09        \\
Ideology         & 0.06         & 0.05         & 0.04   & 0.05        \\
Finance          & 0.40         & 0.39         & 0.38   & 0.42        \\
Shopping         & 0.02         & 0.04         & 0.08   & 0.06        \\
Social \& Gaming & 0.12         & 0.13         & 0.14   & 0.13        \\
Underground      & 0.01         & 0.01         & 0.01   & 0.01        \\
Utility          & 0.08         & 0.11         & 0.08   & 0.11        \\
Others           & 0.13         & 0.15         & 0.12   & 0.09        \\ \bottomrule
\end{tabular}
\caption{Annotation results for bot domain categorization (i.e., the distribution of each category).}
\end{table} 
\end{document}